# Terrestrial planet formation from lost inner solar system material

Christoph Burkhardt[1]*, Fridolin Spitzer[1], Alessandro Morbidelli[2], Gerrit Budde[3], Jan H. Render[1]†, Thomas S. Kruijer[4,5], Thorsten Kleine[1,6]

[1]Institut für Planetologie, University of Münster, Wilhelm-Klemm-Str. 10, 48149 Münster, Germany.
[2]Laboratoire Lagrange, UMR7293, Université de Nice Sophia-Antipolis, CNRS, Observatoire de la Côte d'Azur, Boulevard de l'Observatoire, 06304 Nice Cedex 4, France.
[3]Division of Geological and Planetary Sciences, California Institute of Technology, 1200 E California Blvd, Pasadena, CA 91125, USA.
[4]Museum für Naturkunde, Leibniz Institute for Evolution and Biodiversity Science, Invalidenstraße 43, 10115 Berlin, Germany.
[5]Institut für Geologische Wissenschaften, Freie Universität Berlin, Malteserstraße 74-100, 12249 Berlin, Germany.
[6]Max Planck Institute for Solar System Research, Justus-von-Liebig-Weg 3, 37077 Göttingen, Germany.

*Correspondence to: burkhardt@uni-muenster.de

†Present address: Nuclear and Chemical Sciences Division, Lawrence Livermore National Laboratory, 7000 East Ave., Livermore, CA 94550, USA.

## Abstract

Two fundamentally different processes of rocky planet formation exist, but it is unclear which one built the terrestrial planets of the solar system. Either they formed by collisions among planetary embryos from the inner solar system, or by accreting sunward-drifting millimeter-sized 'pebbles' from the outer solar system. We show that the isotopic compositions of Earth and Mars are governed by two-component mixing among inner solar system materials, including material from the innermost disk unsampled by meteorites, whereas the contribution of outer solar system material is limited to a few percent by mass. This refutes a pebble accretion origin of the terrestrial planets, but is consistent with collisional growth from inner solar system embryos. The low fraction of outer solar system material in Earth and Mars indicates the presence of a persistent dust-drift barrier in the disk, highlighting the specific pathway of rocky planet formation in the solar system.

## Introduction

Rocky planets may have formed by two fundamentally different processes (Fig. 1). In the classic model of oligarchic growth, the accretion of Moon- to Mars-sized planetary embryos in the protoplanetary disk of gas and planetesimals was followed, after gas removal, by a protracted phase of mutual impacts among the embryos, leading to the final assembly of the terrestrial planets (*1*). More recently, an alternative model has been proposed, in which planets grow by accreting "pebbles" from the outer solar system, which drift sunwards through the disk due to gas drag (*2, 3*). Pebble accretion is very effective at forming giant planet cores (*4, 5*), and may have also formed the terrestrial planets of the solar system (*6, 7*). Determining which of these two processes governed the formation of the terrestrial planets of the solar system is crucial for understanding the solar system's architecture and dynamical evolution, and for placing planet formation in the solar system into the context of general planet formation processes, such as those observed in exoplanetary systems.

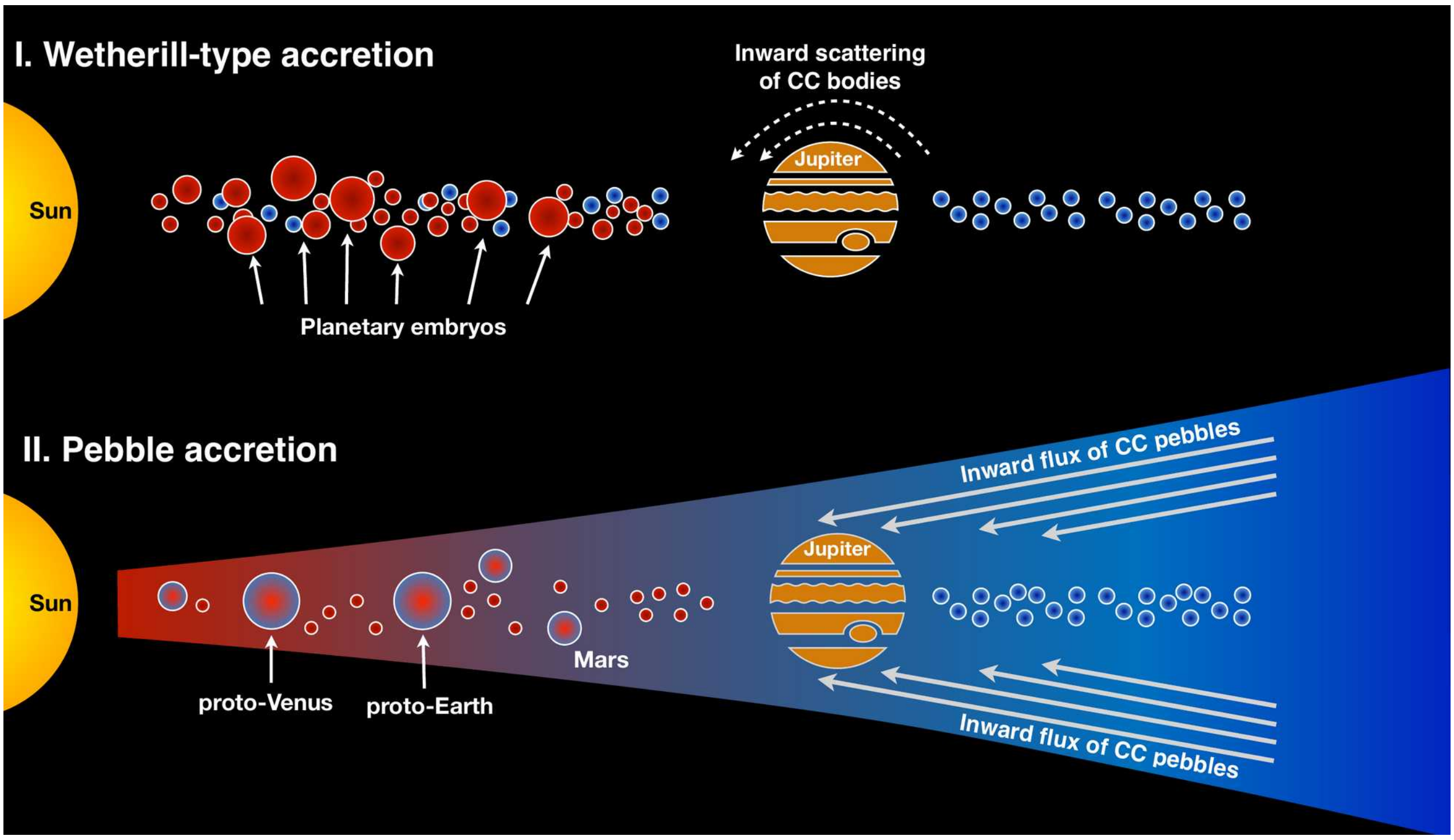

**Fig. 1. Possible scenarios of terrestrial planet formation.** In the classic 'Wetherill-type' model of oligarchic growth, the terrestrial planets formed by mutual collisions among Moon- to Mars-sized planetary embryos after the gas disk dissipated, and accreted only a small fraction of CC planetesimals, which were scattered inwards during Jupiter's growth and/or putative migration. Alternatively, the terrestrial planets may have formed within the lifetime of the gas disk by efficiently accreting "pebbles" from the outer solar system, which drift sunwards through the disk due to gas drag. The two models differ in the amount of outer solar system (CC) material accreted by the terrestrial planets, which may be quantified using nucleosynthetic isotope anomalies.

Oligarchic growth and pebble accretion are both dynamically possible, and so they cannot be distinguished solely on theoretical grounds. Instead, the answer may be sought in the origin of the accreted material, because the two accretion models predict vastly different amounts of outer solar system material incorporated into the terrestrial planets. In the oligarchic growth model, the terrestrial planets predominantly accreted from embryos from the inner solar system, with only a few percent of a planet's mass coming from outer solar system bodies (*8*). By contrast, in the pebble accretion model, much of the terrestrial planets accreted from dust and pebbles drifting inwards from the outer solar system, amounting to ~30-50% by mass of outer solar system material in Earth and Mars (*7, 9, 10*). Thus, the amount of outer solar system material accreted by the terrestrial planets is the key discriminant between the oligarchic growth and pebble accretion models (Fig. 1).
The amount of outer solar system material accreted by the terrestrial planets may be determined using nucleosynthetic isotope anomalies. These arise through the heterogeneous distribution of presolar matter within the solar protoplanetary disk and provide a record of the heritage of a planet's building material (*11-14*). In particular, these isotope anomalies can distinguish between non-carbonaceous (NC) and carbonaceous (CC) meteorites, which most likely represent the inner and outer solar system, respectively (*11*). Thus, the isotopic compositions of Earth and Mars with respect to the NC and CC reservoirs provide a pathway for determining how much outer solar system material was accreted by these bodies. However, whereas some studies concluded that Earth and Mars largely (>95%) accreted from inner solar system (NC) material (*13, 15, 16*), which would be consistent with the oligarchic growth model, others argued that Earth and Mars accreted significant amounts (~30–50%) of CC material (*9, 10*), which in turn would be consistent with a pebble accretion origin (*7*). As such, current interpretations of the isotopic data cannot distinguish between the two models of terrestrial planet formation, because both low and high amounts of outer solar system material in the terrestrial planets have been inferred.

These disparate conclusions reflect that previous studies either considered isotope anomalies of only a subset of elements (*9, 10*), or made specific assumptions about which combination of known meteorites best represents the precursor material of the planets (*13*). Moreover, these studies neglected the observation that Earth likely incorporated material that remained unsampled among meteorites (*12, 17-19*), in which case the amount of CC material in Earth cannot be determined by assuming that meteorites represent its building material. Consequently, accurately determining the bulk fraction of outer solar system material in the terrestrial planets in a self-consistent manner requires not only that the isotope anomalies of a representative set of elements is considered, but also an assessment of the full isotopic variability among the precursor objects of the terrestrial planets, including the planetary building material that remained unsampled among meteorites.

Here we use the recent observation of correlated isotope variations among NC meteorites (*20*) to show that both Earth and Mars incorporated material unsampled among meteorites, determine the provenance and isotopic composition of this lost planetary building material, and use this information to assess the amount of CC material accreted by Earth and Mars. To this end, we also report new high-precision Ti, Zr, and Mo isotopic data for martian meteorites, which are crucial for better defining the isotopic composition of Mars with respect to NC and CC meteorites and, therefore, for determining the fraction of outer solar system material accreted by Mars.

## Results

### *Isotope composition of Mars*

The isotopic compositions of Ti, Zr, and Mo in Mars are poorly defined because until now only a few samples have been analysed and because the precision of prior data is insufficient to unambiguously resolve the small isotopic shifts that may arise through the incorporation of CC material into Mars. To resolve these issues, we obtained Ti, Zr, and Mo isotope data for a comprehensive set of martian meteorites, including samples from the major geochemical reservoirs on Mars as sampled by meteorites (Materials and Methods). As such, the new isotopic data of this study provide a good estimate for the isotopic composition of bulk silicate Mars (BSM) as sampled by meteorites. All isotopic data are reported in the ε-notation as parts-per-ten thousand deviations from terrestrial standard values (see Materials and Methods).

The ten martian meteorites analyzed in this study have indistinguishable Ti isotopic compositions (Supplementary Materials Table S1), with a mean $\varepsilon^{50}$Ti anomaly of –0.42±0.05 (95% conf.). This value is consistent with, but more precise than the mean $\varepsilon^{50}$Ti values calculated from martian meteorite data reported in prior studies (*21-23*), where some individual samples indicated scatter outside the quoted uncertainties (Supplementary Materials Figure S1). Our results corroborate earlier conclusions that the Ti isotopic composition of Mars is intermediate between enstatite and ordinary chondrites, which belong to the NC group of meteorites (Fig. 2, Supplementary Materials Fig. S1). The Zr isotope composition of Mars has so far not been precisely determined. To this end, we report high-precision Zr isotope data for six martian meteorites, which all exhibit indistinguishable Zr isotope anomalies (Supplementary Materials Table S2) with a mean $\varepsilon^{96}$Zr of 0.28±0.03 (95% conf.) (Fig. 2, Supplementary Materials Fig. S2). As for Ti, the Zr isotopic composition for Mars is intermediate between enstatite and ordinary chondrites and also overlaps with the composition of HED (howardite-eucrite-diogenite) meteorites (Supplementary Materials Fig. S2).

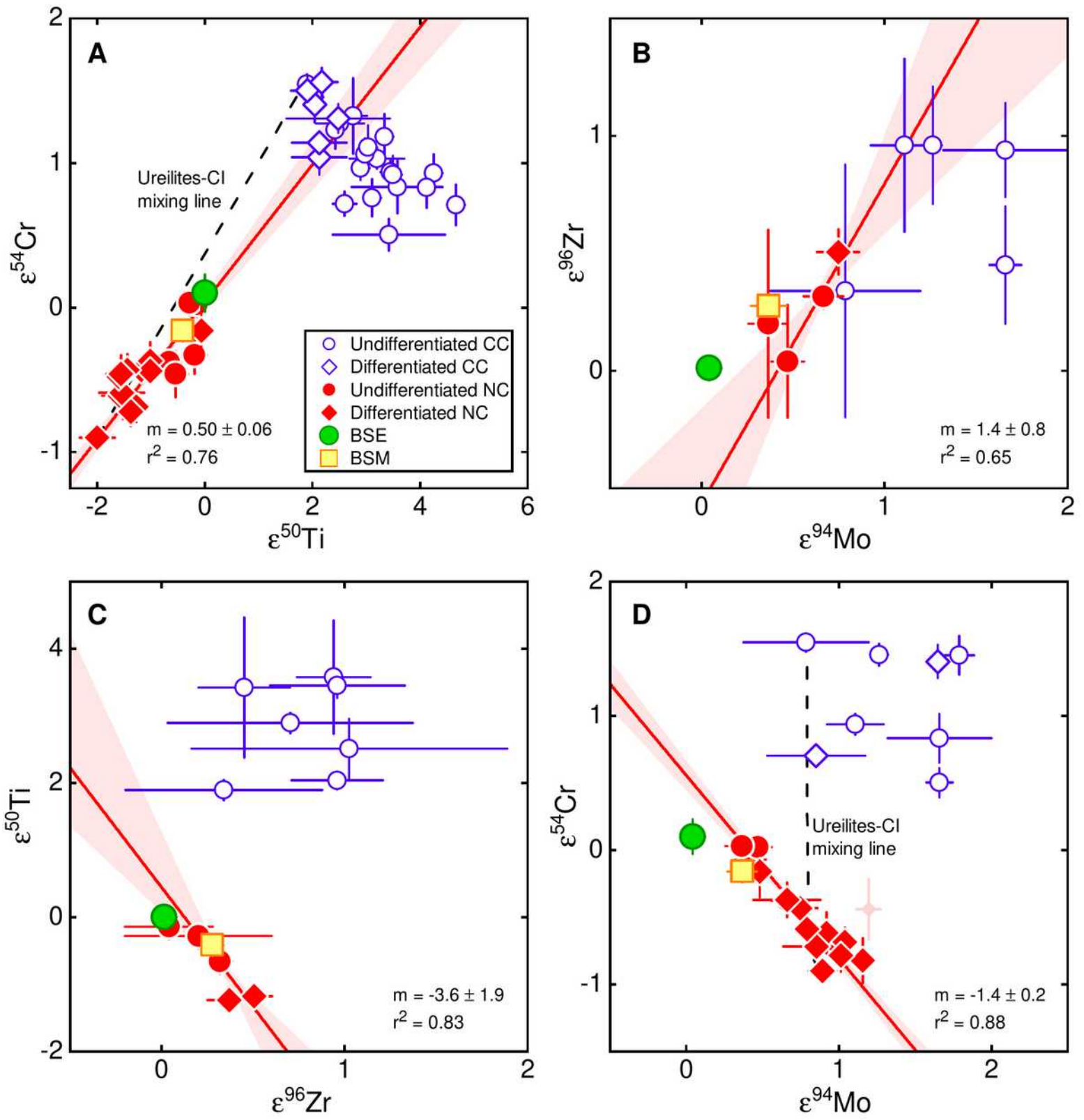

**Fig. 2. Isotope anomalies of meteorites and silicate portions of Earth (BSE) and Mars (BSM).** NC meteorites representing the inner solar system, are shown in red, CC meteorites, representing the outer solar system are shown in blue. For all element pairs, the isotope anomalies among NC meteorites are linearly correlated, with the BSE plotting at one end of these correlations. Solid red lines indicate linear regressions for NC data with corresponding 1σ error envelopes. (**A**) Supernova-associated isotope anomalies in $\varepsilon^{50}$Ti and $\varepsilon^{54}$Cr are positively correlated among NC meteorites. The BSE and BSM plot at the upper end of this correlation pointing towards CC meteorites. (**B**) Anomalies in $\varepsilon^{96}$Zr and $\varepsilon^{94}$Mo are also correlated among NC meteorites, but the BSE plots at the end of the NC meteorite trend pointing away from CC meteorites, towards compositions enriched in *s*-process nuclides. (**C,D**) The BSE also plots at the end of NC correlations for anomalies in Fe-group elements ($\varepsilon^{50}$Ti and $\varepsilon^{54}$Cr) and heavier elements, again pointing away from CC meteorites. Dashed line in (**A**) indicates proposed mixing scenario between ureilite-like and CI-like materials (*9*, *10*), but this mixture cannot account for the compositions of BSE and BSM in multi-dimensional isotope space (**D**). The relationships exemplified here are valid for anomalies in all elements (Fig. 5). Brachinites (light red diamond in panel D) show decoupled Mo–Ru isotope systematics (*52*) and were excluded from the regression. Data and data sources are given in Supplementary Materials. Error bars indicate 95% confidence intervals.

We also report results of eight high-precision Mo isotope measurements on a total of ~22 g of martian material, representing 17 different meteorites (Materials and Methods, Supplementary Materials Table S3). Owing to the low Mo contents of martian meteorites, and the limited mass available for some of the samples, the Mo from some samples has been combined for a single Mo isotope analysis (see Material and Methods). This is justified because any isotope heterogeneity inherited from Mars' building blocks has likely been homogenized during martian differentiation, as is evident from the homogeneous isotope composition among individual martian meteorites observed for other elements like Ti and Zr. Consistent with this, all martian meteorites analyzed in this study have identical Mo isotope anomalies within the external reproducibility (2 s.d.) of the measurements. In a diagram of $\varepsilon^{95}$Mo versus $\varepsilon^{94}$Mo, the martian meteorites plot between the NC- and CC-lines (Fig. 3), indicating that the BSM's Mo derives from both the NC and CC reservoirs. Thus, unlike for Ti and Zr, which reveal no unambiguous evidence for the presence of CC material in Mars, the Mo isotopic data indicate that Mars accreted some CC material. The fraction of CC-derived Mo, as calculated using the lever rule from the BSM's position between the NC- and CC-lines, is 0.4±0.3 (2 s.d.), which is similar to the value of 0.46±0.15 calculated for the BSE using the same approach (*18*). The calculation of CC-derived Mo mass fractions in this manner is strictly valid only if the NC- and CC-lines are parallel, but recent work has shown that the NC-line has a slightly shallower slope than the CC-line (*20*). Nevertheless, the effect of these slightly different slopes on the calculated CC-derived Mo mass fractions in the BSM and BSE is less than 0.1 and, therefore, smaller than the overall uncertainty on these mass fractions. Finally, as a moderately siderophile element, Mo in BSM records only the last ~20% of its accretion (Materials and Methods), and so the fraction of CC material accreted by Mars during its entire accretion history cannot be determined from the Mo isotopic data alone, but requires considering the isotopic composition of Mars in multi-element isotope space (see below).

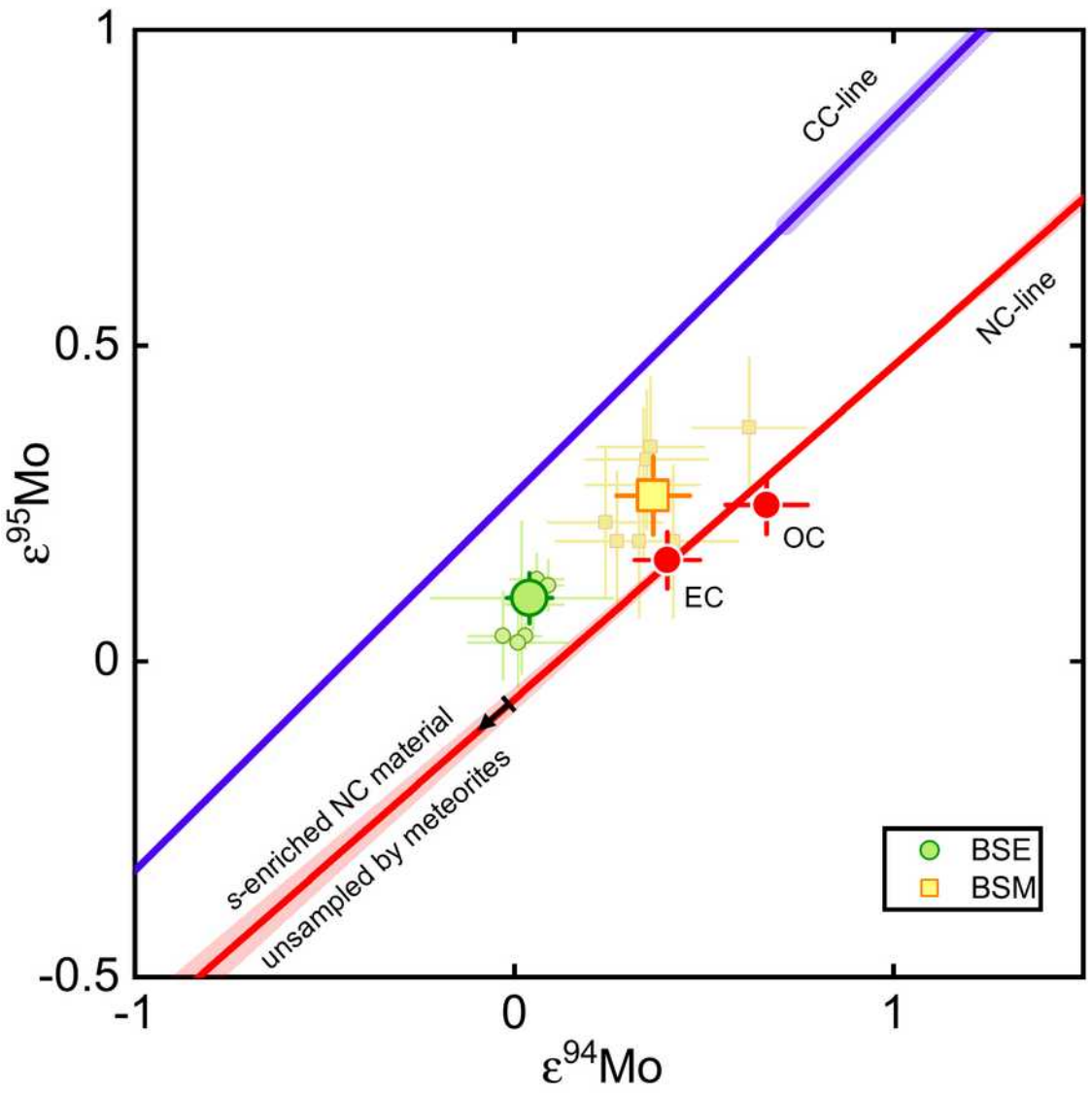

**Fig. 3. Anomalies in $\varepsilon^{95}$Mo and $\varepsilon^{94}$Mo for bulk silicate Earth and bulk silicate Mars.** NC and CC meteorites plot on two approximately parallel lines, the NC- and CC-lines (*18, 20*). The isotopic variations along these lines predominantly reflect *s*-process variations, whereas the offset between them reflects an *r*-process excess in the CC over the NC reservoir. Solid red line indicates linear regression for NC data with corresponding 1σ error envelope (*20*). Both, the BSE and BSM as defined by averaging Mo isotope data (small symbols) of terrestrial mantle derived rocks (*18*) and martian meteorites plot between the lines, indicating the accretion of some CC material to both Earth and Mars. Also shown is the position of the unsampled, *s*-process-enriched material from the inner solar system (starting at arrow). The composition of the BSE and the BSM can be reproduced as a mixture between this component and known NC and CC meteorites (blue shaded area on CC line). Error bars indicate 95% confidence intervals.

### *Building material of Earth and Mars*

The new isotopic data obtained in this study enable direct comparison of the isotopic compositions of bulk silicate Earth (BSE), BSM, and meteorites. A key observation of this comparison is that the BSE plots at one end of multi-element isotope correlations defined by NC meteorites, while the BSM plots close to the compositions of enstatite and ordinary chondrites (Fig. 2). These correlations include isotope anomalies of different nucleosynthetic origin (i.e., supernova- and AGB star-derived nuclides) and comprise elements with different geo- and cosmochemical behaviour, indicating that the isotope variability among NC meteorites reflects mixing between two isotopically distinct components (*20*). Moreover, the NC isotope correlations reveal a striking difference between the Fe-group (e.g., Ca, Ti, Cr, Ni) and heavier elements (e.g., Zr, Mo, Ru, Nd). For the former, the isotope anomalies are positively correlated among NC meteorites, and BSE and BSM plot at the end of this correlation pointing towards CC meteorites (Fig. 2A). Thus, for these elements the isotopic variations among NC meteorites, Earth, and Mars are readily described as mixtures of known NC and CC materials (*10, 11, 13*). However, plotting the isotope anomalies of the heavier elements against each other (Fig. 2B) or against those of the Fe-group elements (Fig. 2C, D) reveals that the BSE plots at the end of the NC meteorite trend pointing away from CC meteorites towards lower $\varepsilon^{94}$Mo and $\varepsilon^{96}$Zr, which represent compositions enriched in nuclides produced in the slow neutron capture process (*s*-process), which takes place in AGB stars. Note that although Earth is isotopically always close to enstatite chondrites, its isotopic composition is distinct and plots beyond the range of compositions recorded in meteorites. This corroborates earlier conclusions that Earth incorporated material that is unsampled among meteorites and, compared to known NC meteorites, is enriched in *s*-process matter (e.g., lower $\varepsilon^{94}$Mo and $\varepsilon^{96}$Zr) (*12, 17, 18, 24*). However, the multi-element isotope correlations among NC meteorites reveal that this unsampled material is also enriched in supernova-derived isotopes of some Fe-group elements (e.g., $\varepsilon^{50}$Ti, $\varepsilon^{54}$Cr). This observation holds for all elements showing well-resolved nucleosynthetic isotope variations among NC meteorites (Fig. 4) and, therefore, is a ubiquitous feature of the building material of the terrestrial planets in the solar system.

The observation that BSE and BSM plot on or close to the NC isotope correlation lines suggests that Earth and Mars predominantly accreted from NC material whose compositions followed the same isotopic gradient as the NC meteorites, and at least for Earth include the unsampled *s*-process enriched component (Fig. 2B-D). By contrast, CC bodies from the outer solar system do not seem to have been a major contributor to Earth and Mars, because otherwise the BSE and BSM would plot off the NC isotope trends towards the composition of the CC reservoir. Nevertheless, some CC addition to Earth and Mars is necessary to account for the observation that the BSE and BSM plot between the NC- and CC-lines in the $\varepsilon^{95}$Mo–$\varepsilon^{94}$Mo diagram (Fig. 3). To assess these observations more quantitatively, we have developed a mixing model that reduces the isotopic composition of a planet's silicate mantle (i.e., BSE or BSM) to the contribution of three main components: (1) NC material contributing to the Mo in the planet's present-day mantle, (2) NC material not contributing to the Mo in the planet's present-day mantle, and (3) CC material (see Material and Methods). The distinction between two NC components is necessary, because the mantle's isotopic composition for a siderophile element like Mo predominantly records the later stages of accretion (*13, 18*). With this model, the fraction of the mantle's Mo delivered by NC bodies, and the Mo isotopic composition of these bodies can be determined from the BSE's or BSM's $\varepsilon^{94}$Mo and $\varepsilon^{95}$Mo (*18*), the $\varepsilon^{95}$Mo–$\varepsilon^{94}$Mo relationship among NC meteorites (*20*), and the average $\varepsilon^{94}$Mo and $\varepsilon^{95}$Mo of CC meteorites (*25*). As the isotopic compositions of both NC components for all other elements depend on the respective values of $\varepsilon^{94}$Mo via the NC correlation lines, the bulk fraction of CC material in Earth and Mars can be determined by using the isotope correlations of two lithophile elements (that is, elements unaffected by core formation), $\varepsilon^{54}$Cr (or $\varepsilon^{50}$Ti) and $\varepsilon^{96}$Zr, versus $\varepsilon^{94}$Mo (Fig. 2B, D). Finally, a Monte Carlo approach was used to account for the associated uncertainties on all input quantities (e.g., the slope of the NC correlation lines or the isotopic variations among CC meteorites), and to calculate probability distribution functions (PDF) for (1) the fraction of the mantle's Mo delivered by NC bodies, (2) the Mo isotopic composition of these bodies, and (3) the bulk CC fraction accreted by Earth and Mars (see Material and Methods).

The model reveals that about 70% of the BSE's Mo was delivered by NC material with an average isotopic composition peaking at an $\varepsilon^{94}$Mo value of about –0.6 (Fig. 5A, B), indicating that most of the BSE's Mo derives from NC objects that, on average, are distinct from known meteorites, all of which are characterized by positive $\varepsilon^{94}$Mo. This does not mean that all NC material accreted by Earth had this particular $\varepsilon^{94}$Mo, but rather that the NC accretionary mix probably included objects with more positive (such as NC meteorites) and more negative $\varepsilon^{94}$Mo (the unsampled material from above). In fact, the offset of the BSE from the $\varepsilon^{94}$Mo–$\varepsilon^{54}$Cr correlation line (Fig. 2D) indicates that later accreted objects had a composition further up on the correlation line than earlier accreted objects. This is because a siderophile element like Mo is strongly depleted in proto-Earth's mantle due to core formation, and so for siderophile-lithophile element pairs mixing lines between two isotopically distinct impactors are curved. Late impactors, therefore, had a stronger effect on the BSE's Mo than on its Cr isotopic composition, and so the offset of the BSE to the left of the $\varepsilon^{94}$Mo–$\varepsilon^{54}$Cr correlation line implies that late impactors had a composition further up on the line.

The results for Mars are very similar to those for Earth, but the average $\varepsilon^{94}$Mo of the NC material that contributed to the BSM's Mo is shifted towards slightly more positive values peaking at a value of ~0 (Fig. 5A). Moreover, unlike the BSE, Mars is not clearly offset from the $\varepsilon^{94}$Mo–$\varepsilon^{54}$Cr correlation line (Fig. 2D).

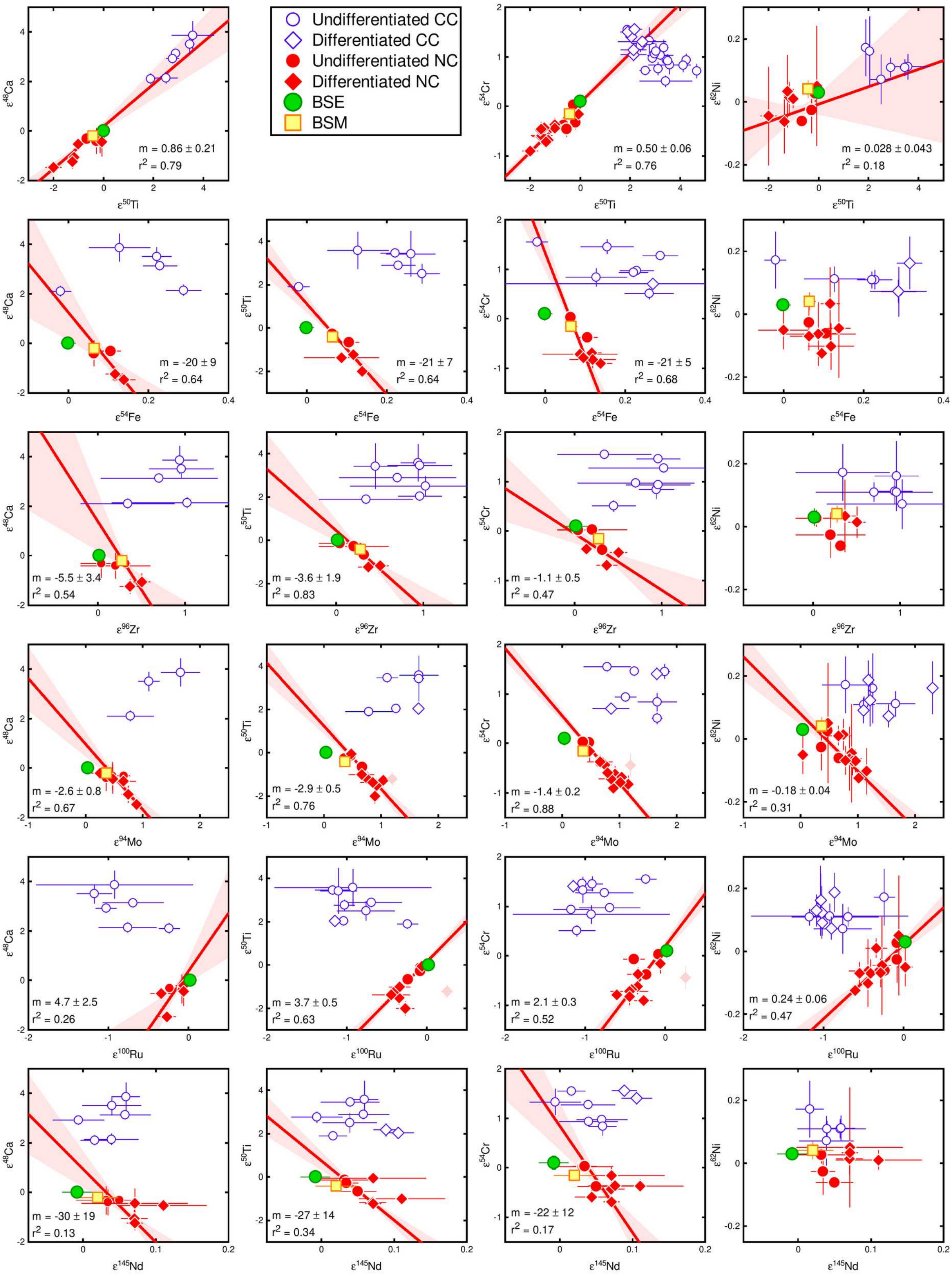

**Fig. 4. Isotope anomalies of NC and CC meteorites, and silicate portions of Earth and Mars in multi-element space.** NC meteorites, representing the inner solar system, are shown in red, CC meteorites, representing the outer solar system are shown in blue. The isotope anomalies among NC meteorites are linearly correlated, with the BSE always plotting at one end of these correlations. Solid red lines indicate linear regression (York-fit) of isotopic data for NC meteorites with corresponding 1σ error envelopes. Brachinite meteorites have decoupled Mo and Ru isotope systematics (*52*) and were excluded from regressions in $\varepsilon^{50}$Ti and $\varepsilon^{54}$Cr vs. $\varepsilon^{94}$Mo and $\varepsilon^{100}$Ru space (light red diamonds). Limited data, data precision, or data spread do currently not allow for meaningful NC regressions in $\varepsilon^{62}$Ni-$\varepsilon^{54}$Fe, $\varepsilon^{62}$Ni-$\varepsilon^{96}$Zr and $\varepsilon^{62}$Ni-$\varepsilon^{145}$Nd space. Data are given in the Supplementary Materials Data S1. Uncertainties represent 95% confidence intervals.

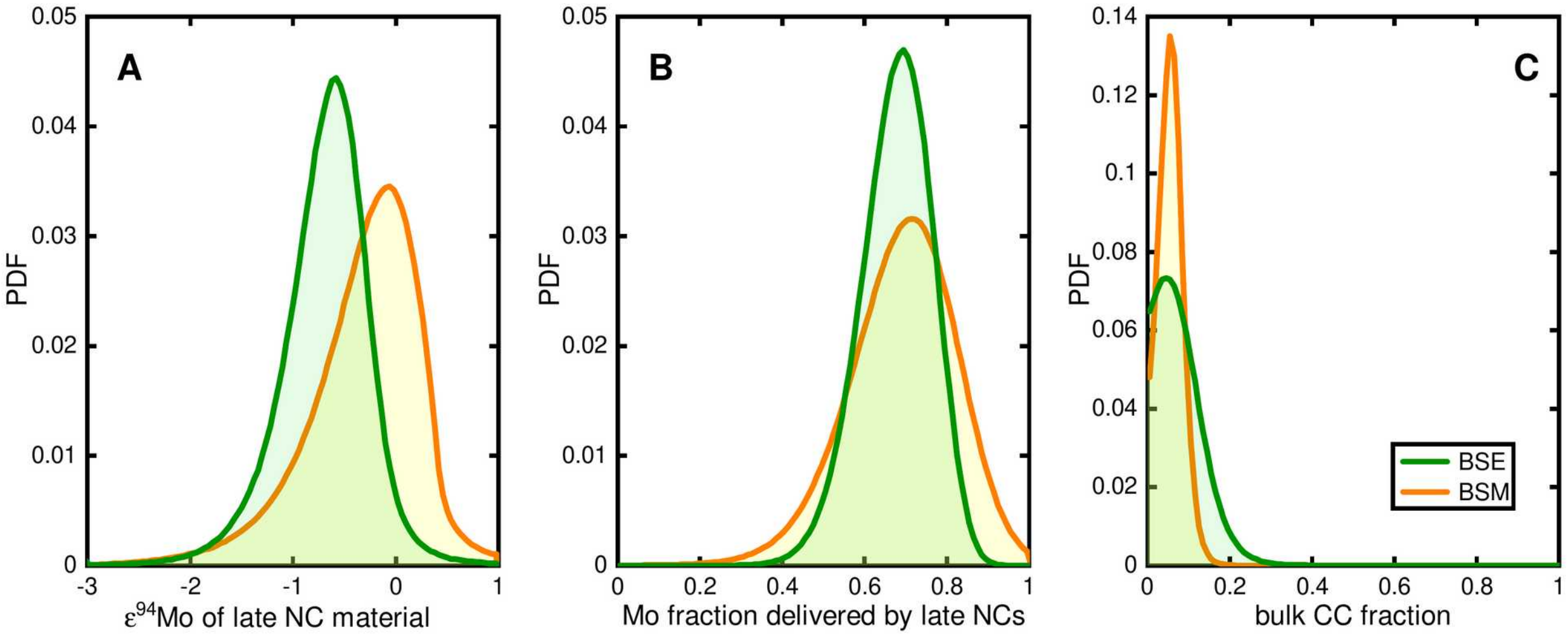

**Fig. 5. Results of Monte Carlo simulations for reproducing the isotopic composition of Earth and Mars in multi-dimensional isotope space.** (**A**) Probability distribution function (PDF) of average $\varepsilon^{94}$Mo delivering NC Mo to the BSE and BSM. (**B**) PDF of fraction of BSE and BSM Mo delivered by NC (the complement being from CC). (**C**) PDF of fraction of bulk Earth and Mars delivered by CC material, with a peak at 4% for both planets.

A key result of our model is that we infer only a small fraction of CC material (with a peak at ~4%) for Earth and Mars (Fig. 5C). Note that the larger CC fractions of ~30–40% inferred by considering solely the Mo isotopic data is not representative for the bulk fraction of CC material in Earth and Mars, but only refers to the last ~10–20% of accretion (see above). Instead, the bulk CC fraction in both planets can only be determined by considering isotopic data for all elements together, including lithophile elements. Our finding that Earth and Mars incorporated only a small bulk fraction of CC material has far-reaching implications for understanding the fundamental process of terrestrial planet formation in the solar system.

## Discussion

### *Lost planetary building material from the inner solar system*

The low mass fraction (~4 %) of outer solar system in Earth and Mars determined in this study is in stark contrast to prior proposals that the isotopic composition of Earth and Mars reflect ~1:1 and ~2:1 mixtures of inner disk material having an ureilite-like isotopic composition with inward-drifting CI-like dust from the outer solar system (*9*, *10*). This large fraction of CC material was inferred solely based on isotope anomalies in Fe-group elements for which, as noted above, BSE and BSM plot at the end of the NC isotope correlation that points towards the CC compositional field (Fig. 2A). However, in multi-element isotope space, mixtures between ureilites and CI chondrites do not pass through the compositions of BSE and BSM, which instead plot furthest away from these mixing lines (Fig. 2C, D). Thus, the accretion of large amounts of CC material to Earth and Mars would only be possible if this material was enriched in *s*-process matter (i.e., negative $\varepsilon^{94}$Mo and $\varepsilon^{96}$Zr) and, in this case, would represent the aforementioned planetary building material that remained unsampled among meteorites. The necessary enrichment of CC material in *s*-process matter may perhaps be possible as a result of filtering (*26*, *27*) or thermal processing (*28*) of CC dust during its transport into the inner solar system.

However, several lines of evidence indicate that the unsampled planetary building material is not CC dust. First, the isotopic composition of the putative *s*-process-enriched CC dust may be quantified by extrapolation of the NC-line to the intersection with the typical Ca, Ti, Cr, and Ni isotopic composition of CC meteorites. This is possible because the isotopes of Ca, Ti, Cr, and Ni are not produced in significant amount in the *s*-process, such that processing of *s*-process carriers would not significantly change the isotopic composition of these elements. However, the projected isotope compositions determined in this manner plot on the NC-line in Mo isotope space and not, as would be required, at the intersection of the NC- and CC-lines (Supplementary Materials Fig. S3). Second, if the isotopic trend among NC meteorites were to reflect the progressive addition of CC dust, then the isotopic compositions of NC bodies would have evolved over time towards more CC-like compositions. However, there is no obvious correlation between isotope anomalies and accretion ages of NC meteorite parent bodies, indicating that the isotope variability among NC meteorites does not reflect a temporal evolution of inner disk composition (*20*). Third, the isotope variability among NC meteorites seems to follow a heliocentric gradient. Although their original formation location is not well known, it is often assumed that enstatite chondrites, which have the lowest $\varepsilon^{94}$Mo and $\varepsilon^{96}$Zr and highest $\varepsilon^{50}$Ti and $\varepsilon^{54}$Cr among NC meteorites, formed closer to the Sun than other NC meteorites such as ordinary chondrites or ureilites (*29*). This gradient is also consistent with the higher $\varepsilon^{94}$Mo and $\varepsilon^{96}$Zr and lower $\varepsilon^{50}$Ti and $\varepsilon^{54}$Cr of Mars compared to Earth, and would also be consistent with the lower $\varepsilon^{50}$Ti and $\varepsilon^{54}$Cr of HED meteorites, provided that the HED parent body, asteroid Vesta, formed close to its present-day position in the asteroid belt (Fig. 2). Together, these observations suggest that the unsampled *s*-process-enriched material derives from the innermost disk sunward of the formation location of known NC meteorites. This material likely remained unsampled among meteorites because of the low probability for scattering objects from the innermost disk into the asteroid belt, where the parent bodies of meteorites remained stored over the age of the solar system.

***Origin of isotopic difference between Earth and Mars***

The average $\varepsilon^{94}$Mo of the NC material contributing to Mo in BSE and BSM is more negative for Earth than for Mars, indicating that Earth accreted more of the unsampled, *s*-process-enriched material from the innermost solar system than Mars did. This is consistent with Mars' formation at greater heliocentric distance than Earth and with our model that material closer to the Sun plots on the multi-element NC correlation lines beyond the terrestrial composition. The isotopic difference between Earth and Mars, therefore, likely reflects their formation at different radial locations in the disk. This resulted in different ratios of objects with more positive (such as NC meteorites) and more negative $\varepsilon^{94}$Mo (the unsampled material from above) in the NC accretionary mixtures of Earth and Mars, which in turn are reflected in different positions along the NC isotope correlation.

Formation of Earth and Mars from heterogeneous mixtures of NC bodies seems at odds with the observation that for *all* elements investigated so far Earth is isotopically very similar to enstatite chondrites (*13*), and that Mars is typically intermediate between enstatite and ordinary chondrites (*15, 16*). This has been interpreted to reflect accretion of Earth and Mars from isotopically homogeneous materials (*13, 15*). However, the isotopic link between Earth and enstatite chondrites arises naturally from the multi-element isotope correlations among NC meteorites. In fact, once the heterogeneous mixture that made Earth is, by chance, isotopically similar to a given meteorite group for one element, the same isotopic similarity will necessarily hold for all other elements. The same applies to Mars, but at a different position on the NC isotope correlation.

***Mode of terrestrial planet formation in the solar system***

A key result of our study is that once a representative set of elements is considered, the isotopic composition of Earth and Mars indicate that both planets accreted only a small fraction of CC material, with a peak at 4% by mass (Fig. 5C). Thus, contrary to some recent studies (*7, 9, 10*), we conclude that the terrestrial planets cannot have formed by accretion of large masses (~40%) of CC pebbles from the outer solar system. Instead, the terrestrial planets could only have formed by pebble accretion if the vast majority of the pebbles had an NC isotopic composition. To this end, it has been suggested that CC material initially resided far from the Sun, and that it reached the terrestrial planet region by radial drift only after some time *t*, before which the planets grew by accreting NC pebbles (*7*). Within this framework, the fraction of CC material in the terrestrial planets would be higher the earlier the change in pebble composition from NC to CC in the terrestrial planet region would have occurred. For instance, to match the purported high CC fraction in the terrestrial planets (*9, 10*) in this particular model would require this change to have occurred at ~3.8 Ma after solar system formation (*7*). However, the low CC fraction in the terrestrial planets determined in the present study indicates that this assumed change in pebble composition would have to have occurred at ~5 Ma, which is the assumed lifetime of the disk in this model (*7*). Moreover, for a typical pebble Stokes number of 0.001, the dust reaching the terrestrial planet region at 5 Ma would have originally been located at ~100 AU. However, given that the parent bodies of some CC iron meteorites formed within 1 Ma after the start of the solar system (*30*), and considering that these bodies have been implanted into the asteroid belt during the growth of the giant planets (*8, 31*), it is implausible that the original boundary between the NC and CC reservoirs was located as far out as ~100 AU. Consequently, the most reasonable and straightforward interpretation is that the low fraction of CC material in the terrestrial planets implies a dramatically reduced influx of dust into the inner disk. It has been suggested that a pressure maximum in the disk or the snow line acted as barriers against grain drift and, hence, were responsible for the initial separation of the NC and CC reservoirs (*32-34*). However, it is unclear as to whether these transient disk structures were capable of maintaining an efficient separation of the NC and CC reservoirs over the entire lifetime of the disk. Hence, we propose that the most plausible cause for the long-lasting separation of the NC and CC reservoirs is the formation of Jupiter, which has been shown to act as an efficient and persistent barrier against the inward drift of dust and pebbles (*26, 27, 35*).

In spite of the low bulk CC fraction, Earth and Mars must have accreted some CC bodies during later growth stages, because otherwise the BSE and BSM would not plot between the NC- and CC-lines (Fig. 3). The CC contribution to the BSE's and BSM's Mo are remarkably similar (Fig. 5B), indicating a common process of CC accretion for both bodies. For instance, although the mixed NC-CC Mo isotopic composition of the BSE may reflect delivery of CC material by the Moon-forming impactor (*18*), a similar process did not occur for Mars. Moreover, our results indicate that CC planetesimals were already present in the inner disk when Mars reached its final mass at ~5 Ma after solar system formation (*36*). We, therefore, suggest that Earth and Mars, as well as other planetary embryos in the inner solar system, most likely accreted CC material through planetesimals, which contributed little overall mass but nevertheless left a significant imprint on their mantle's Mo isotopic compositions. This is consistent with chemical models for core-mantle differentiation on Earth, which also argue for the accretion of CC planetesimals towards the end of accretion (*37*). Finally, we note that the estimate of ~4% CC material in Earth and Mars is sufficient to account for the volatile element inventory of both planets, even under the unlikely condition that all NC material accreted by them was volatile-free (*38-40*).

The subordinate role of outer solar system material among the accretionary mix of Earth and Mars is consistent with the classic model of terrestrial planet formation by oligarchic growth through collisions among inner solar system planetesimals and planetary embryos, with only little contamination from outer solar system objects. Of note, our estimate of only ~4% CC material in Earth and Mars is consistent with the results of dynamical models in which CC asteroids are scattered into the inner solar system during the growth (*8*) and/or migration (*41*) of Jupiter. This combined with the efficient isolation of the inner disk from inward-drifting dust from the outer solar system required by the isotopic data bears testimony to the crucial role of Jupiter for determining the fundamental process that built terrestrial planets in the solar system.

## Materials and Methods

### *Samples, sample digestion, and initial chemistry*

The 17 martian meteorites of this study derive from distinct sources of the martian mantle, which were established during the early differentiation of a martian magma ocean. Samples include five depleted shergottites (DaG 476, LAR 12011, NWA 7635, SaU 005, Tissint), three intermediate shergottites (ALH 77005, EETA 79001, NWA 7042), four enriched shergottites (LAR 12011, NWA 4864, RBT 04262, Zagami), three nakhlites (Nakhla, NWA 10153, MIL 03346), orthopyroxenite ALH 84001 and augite basalt NWA 8159. Individual sample masses ranged from 0.2 to 3.6 g, which combined amount to a total of ~22 g of martian material. Following cleaning of fresh interior pieces by ultrasonication in ethanol, samples were crushed in an agate mortar, and sample powders were digested in Savillex PFA beakers with a 2:1 mixture of HF:$HNO_3$ on a hotplate at 130-150°C for 2-5 days. Titanium, Zr, and Mo (this study), as well as Cr (*42*), and Nd and W (*43*) were then separated from the sample matrix using ion exchange chromatography following our previously established procedures (*18, 24, 44-47*). Samples were loaded onto columns filled with AG1X8 (4 ml, 200-400 mesh) in 0.5M HCl-0.5M HF (25 ml). Samples larger than 0.5 g were split over two or more columns. Following elution of matrix elements, Cr and REEs with further addition of 0.5M HCl-0.5M HF (10 ml), Ti, Zr, and Hf were eluted with 10 ml 8M HCl–0.05M HF. Finally, W and Mo were rinsed off the columns in 15 ml 6M HCl-1M HF and 10 ml 3M $HNO_3$, respectively. Two large samples (3.56 g of DaG 476 and 3.42 g of Tissint) were exclusively processed for this study and were digested for 7 days in HF-$HNO_3$ followed by HCl-$HNO_3$. Also, for these two samples each 0.5 g split was loaded in 75 ml of 0.5M HCl-0.5M HF on the column, and the Ti, Zr, Hf elution step was omitted, such that these elements were eluted together with W. For all samples split over two or more columns, the individual matrix (+Cr, REE), Ti-Zr-Hf, Mo, and W cuts from each column were combined after the first column.

***Titanium clean-up chemistry and isotope measurements***
The Ti-Zr-Hf cuts of eight shergottites (DaG 476, LAR 12011, NWA 7635, SaU 005, Tissint, ALH 77005, EETA 79001, NWA 7042), orthopyroxenite ALH84001, and augite basalt NWA 8159 were dried and re-dissolved in 12M $HNO_3$. They were subsequently passed through a two-stage ion exchange procedure involving TODGA (2 ml) and AG1X8 (0.8 ml, 200-400 mesh) ion exchange resins (*48*) to separate Ti from Zr and Hf. Titanium isotope measurements were made on a ThermoScientific Neptune *Plus* MC-ICPMS in high-resolution mode at the Institut für Planetologie in Münster (*45*). Solutions containing about 400 ppb Ti were introduced through a Cetac Aridus II desolvating system, resulting in a total ion beam of ~$4.5\times10^{-10}$ A. Each Ti isotope measurement consisted of two lines of data acquisition. After 30 s baseline integrations (deflected beam), ion beams on all Ti isotopes as well as $^{51}V$ and $^{53}Cr$ were measured in line 1 in blocks of 40 cycles of 4.2 s each, whereas in line 2, ion beams on $^{44}Ca$ and all the Ti isotopes were measured in 20 cycles of 4.2 s integrations each. Instrumental mass bias was corrected using the exponential law and $^{49}Ti/^{47}Ti = 0.749766$. Titanium isotope compositions are reported as $\varepsilon^{i}Ti$ values relative to the mean of bracketing measurements of the OL-Ti (Origins Lab) standard, where $\varepsilon^{i}Ti = [(^{i}Ti/^{47}Ti)_{sample}/(^{i}Ti/^{47}Ti)_{standard} - 1] \times 10^4$. Results were normalized to the mean of the terrestrial standards (BHVO-2, BIR1a, OL-Ti) processed through the chemistry along with the samples. The latter was done because we observed a slight offset in $\varepsilon^{46}Ti$ (-0.08±0.05, 95% confidence interval, N = 33) and $\varepsilon^{50}Ti$ (-0.05±0.04) between the processed terrestrial rock samples and standards and the unprocessed Ol-Ti bracketing standard. Such offsets might be caused by an improper mass fractionation correction assuming the exponential law, or unaccounted matrix effects, and have been observed (and corrected this way) in other high-precision Ti isotope studies before (*49*). The uncertainty introduced by this correction was propagated into the analytical uncertainties of the Ti isotope measurements for all samples by quadratic addition. The external reproducibility (2SD) of the standard measurements during the course of this study was ±0.26 $\varepsilon^{46}Ti$, ±0.20 $\varepsilon^{48}Ti$, and ±0.31 $\varepsilon^{50}Ti$. The Ti isotope compositions of the martian meteorites analysed in this study are provided in the Supplementary Materials Table S1. All 10 samples of this study have indistinguishable Ti isotopic compositions with a mean $\varepsilon^{50}Ti$ of -0.42±0.05, which provides the current best estimate for the Ti isotopic composition of Mars.

***Zirconium clean-up chemistry and isotope measurement***
For samples with sufficiently high Zr contents, the Zr-Hf cuts from the first column of the Ti chemistry were evaporated to dryness and redissolved in 2 ml 3M $HNO_3$-1wt% $H_2O_2$ for further purification following the procedure of ref. (*50*). Sample solutions were loaded onto PFA shrink columns filled with Eichrom LN spec resin (100-150 mesh) for separation of Zr from residual Fe, Ti, and Hf. First, Ti was eluted with 10 ml 3M $HNO_3$-1wt% $H_2O_2$ in 2 ml incremental steps, which was followed by 2 ml 0.28M $HNO_3$ to wash off residual $H_2O_2$ from the resin. After rinsing with additional 10 ml 0.5M $HNO_3$-0.06M HF, Zr was collected in 24 ml 0.5M $HNO_3$-0.06M HF, before Hf was eluted with 6 ml 0.56M $HNO_3$-0.3M HF. The final Zr cuts were treated with concentrated $HNO_3$ and redissolved in 0.5M $HNO_3$-0.01M HF for isotope measurements. The purified Zr cuts had Mo/Zr <$1\times10^{-4}$, Ru/Zr <$1\times10^{-4}$, Ti/Zr <0.1, Fe/Zr <0.01, and Hf/Zr <0.01, all of which are well below the thresholds required for accurate of interference corrections. Yields of the entire separation procedure were >80%. Total procedural blanks were <1ng Zr and were insignificant, given that more than 10 µg Zr was processed for each sample. Zirconium isotope measurements were performed using the Neptune *Plus* MC-ICPMS at the Institut für Planetologie, using Jet sampler and H skimmer cones. Samples were introduced as ~200 ng/g solutions using a Cetac Aridus II desolvator and a Savillex C-flow nebulizer with an uptake rate of 50µl/min, resulting in a total ion beam intensity of 6 to $8\times10^{-10}$ A. Oxide rates were adjusted to <1.5% CeO/Ce and samples and standards were matched in concentration to within <15%. Each measurement consisted of 30 s baseline integrations (deflected beam) followed by 200 cycles Zr isotope ratio measurements of 4.2 s each. Ion beams on all Zr masses (90, 91, 92, 94, and 96) were measured using Faraday cups connected to $10^{11}\Omega$ feedback resistors, and ion beams on masses 95 and 99 were measured using Faraday cups connected to $10^{12}\Omega$ feedback resistors to monitor potential isobaric interferences of Mo and Ru on various Zr isotopes. Instrumental mass bias was corrected by internal normalization to $^{94}Zr/^{90}Zr = 0.3381$ and using the exponential law. The Zr isotope data are reported as $\varepsilon^{i}Zr$ values relative to the mean composition measured for the NIST SRM 3169 Zr standard that was analyzed bracketing the sample measurements, where $\varepsilon^{i}Zr = [(^{i}Zr/^{90}Zr)_{sample}/(^{i}Zr/^{90}Zr)_{standard} - 1] \times 10^4$ . The Zr isotopic composition of two terrestrial rock standards (BHVO-2 and BCR-2) processed along with the samples are indistinguishable from the solution standard. The Zr isotope data of the martian meteorites and the terrestrial basalt standards are provided in the Supplementary Materials Table S2. All martian meteorites have indistinguishable Zr isotope anomalies with a mean $\varepsilon^{96}Zr = 0.28\pm0.03$. This is the first report of Zr isotope data for martian meteorites and as such is the current best estimate for the Zr isotopic composition of Mars.

***Molybdenum clean-up chemistry and isotope measurement***

Except for Tissint, the Mo cuts of the 17 individual sample digestions common to this study and ref. (*43*) did not contain enough Mo (~100 ng) for individual high-precision Mo isotope measurements. The Mo cuts of these samples were, therefore, combined to obtain a total of four samples (Tissint-1, Mars-A, Mars-B, and Mars-C). Combining the Mo cuts from different martian samples is possible because no nucleosynthetic isotope heterogeneity is expected among different martian meteorites. This is because any potential heterogeneity resulting from accretion of isotopically heterogeneous building blocks should have been erased by subsequent melting of the martian mantle and homogenization within a martian magma ocean. This is consistent with the lack of resolved Zr and Ti isotope variations among individual martian meteorites derived from distinct mantle sources observed in this study (see above). Mars-A consists of the combined Mo cuts of the depleted shergottites SaU 005 and DaG 476, Mars-B consists of the combined Mo cuts of the intermediate shergottites ALH 77005, EETA 79001, and NWA 7042, and Mars-C consists of the combined Mo cuts of the remaining 11 samples of the Nd-W isotope study of ref. (*43*) (NWA 8159, RBT 04262, NWA 4864, LAR 12011, LAR 12095, ALH 84001, NWA 7635, Zagami, Nakhla, NWA 10153, MIL 03346). Together with the larger DaG 476 and Tissint samples exclusively digested for this study, this resulted in a total of six martian Mo samples (Tissint-1, Mars-A, Mars-B, Mars-C, DaG 476, and Tissint-2) on which a Mo clean-up chemistry was performed. The purification of Mo followed our established two-stage ion exchange procedure using AG1X8 (2ml, 100-200mesh) and TRU (1ml) ion exchange resin (*18, 30, 51*). After the ion exchange chemistry, the purified Mo cuts were treated with inverse aqua regia to destroy organic compounds before they were taken up in 0.5M $HNO_3$-0.01M HF for isotope measurement. The Mo yields for the separation were between 65 and 82%, and the procedural blanks ranged from 3 to 16 ng. This corresponds to 3 to 9 % of the Mo recovered from each of sample and, given the overall small Mo isotope anomalies in the samples, had no significant effect on measured Mo isotope compositions. The terrestrial rock standards JA-2 and W-2a, which were processed through the analytical procedure along with the martian samples yielded the terrestrial Mo isotopic composition (reported in ref. (*18*)), testifying to the accuracy of the analytical methods. The Mo isotopic compositions were measured using a ThermoScientific Neptune *Plus* MC-ICP-MS at the Institut für Planetologie. Samples were introduced as 100 ng/g Mo solutions using a Savillex C-Flow PFA nebulizer (uptake rate ~50 µl/min) connected to a Cetac Aridus II desolvator, resulting in total ion beam intensities of $\sim 1.1\times 10^{-10}$ A. Each measurement consisted of 40 on-peak baseline integrations, followed by 100 cycles of isotope ratio measurements of 8.4s each. Instrumental mass bias was corrected by internal normalization to $^{98}Mo/^{96}Mo$ = 1.453173 and using the exponential law. Isobaric interferences of Zr and Ru on Mo masses were corrected by monitoring $^{91}Zr$ and $^{99}Ru$. The maximum corrections were <0.5 ε, and were all well within the range of reliable and accurate interference corrections (*51*). The Mo isotope data are reported as $\varepsilon^{i}Mo$ values relative to the mean of bracketing Alfa Aesar Mo standard measurements, where $\varepsilon^{i}Mo = [(^{i}Mo/^{96}Mo)_{sample}/(^{i}Mo/^{96}Mo)_{standard} - 1] \times 10^{4}$. The external reproducibility (2SD) of the standard measurement during the course of this study was ±0.29 $\varepsilon^{92}Mo$, ±0.25 $\varepsilon^{94}Mo$, ±0.18 $\varepsilon^{95}Mo$, ±0.15 $\varepsilon^{97}Mo$, and ±0.14 $\varepsilon^{100}Mo$. The Mo isotope data of the martian samples are reported in the Supplementary Materials Table S3. Within the external reproducibility of the measurements, all martian meteorites have identical Mo isotope compositions and reveal resolved Mo isotope anomalies relative to the standard and terrestrial samples. The mean Mo isotopic composition of the martian meteorites is consistent with an *s*-process deficit relative to Earth’s mantle, and, given that the samples of this study derive from different portions of the martian mantle, provides the current best estimate for the Mo isotopic composition of bulk silicate Mars (BSM). Although there is some scatter in the individual measurements, in a diagram of $\varepsilon^{95}Mo$ versus $\varepsilon^{94}Mo$ the average BSM composition plots between the NC- and CC-lines (Fig. 3). Together, these data indicate that the BSM's Mo derives from both the NC and CC reservoirs, where the fraction of CC-derived Mo, as calculated from its position between the NC- and CC-lines using the lever rule, is 0.4±0.3 (2σ). As a siderophile element, the BSM's Mo predominantly records the last ~20% of accretion, because the Mo from earlier accretion stages has been removed to the core. Thus, the Mo isotopic data indicate that Mars accreted some CC material during the last ~20% of its growth, but the fraction of CC material accreted by Mars during its entire accretion history cannot be determined in this manner. More precise and robust constraints on how much CC material was accreted by Mars can be obtained by considering its isotopic composition in multi-element space (see below and main text).

***Isotopic mixing model for Earth and Mars***

As discussed in the main text, the building blocks of Earth (or Mars) can be reduced to three main components: average NC material contributing to the BSE's Mo budget (denoted $NC_{Late}$), average NC material not contributing to the BSE's Mo (denoted $NC_{Early}$), and CC material. The indexes “Early” and “Late” reflect the siderophile nature of Mo, which means that the material contributing to the BSE's Mo likely accreted later than the material that did not contribute to the final Mo budget of the BSE (that is, the Mo from these objects was entirely removed to Earth's core). These three components come with 5 unknowns: the $\varepsilon^{94}Mo$ values of $NC_{Late}$ and $NC_{Early}$, denoted $\varepsilon^{94}Mo^{NCLate}$ and $\varepsilon^{94}Mo^{NCEarly}$, the fraction $f_{Mo}^{NCLate}$ of the BSE's Mo delivered by $NC_{Late}$, and the fractions of CC and $NC_{Late}$ in the bulk Earth, $f_{Bulk}^{CC}$ and $f_{Bulk}^{NCLate}$. Note that $f_{Mo}^{NCLate}$ should not be confused with the fraction of Earth’s mass that accreted “late”, but merely refers to the fraction of a mantle's NC-derived Mo that has been delivered late. We also note that because for all isotope anomalies, NC meteorites plot on the NC-line, once $\varepsilon^{94}Mo^{NCLate}$ and $\varepsilon^{94}Mo^{NCEarly}$ are known, their isotope anomalies for all other elements are known as well.

To determine $\varepsilon^{94}Mo^{NCLate}$ and $f_{Mo}^{NCLate}$ we use two equations for $\varepsilon^{94}Mo$ and $\varepsilon^{95}Mo$. These are:

$$\varepsilon^{94}Mo^{NCLate} f_{Mo}^{NCLate} + (1 - f_{Mo}^{NCLate})\, \varepsilon^{94}Mo^{CC} = \varepsilon^{94}Mo^{BSE} \quad (1)$$

$$(a_{95Mo}\,\varepsilon^{94}Mo^{NCLate} + b_{95Mo})\, f_{Mo}^{NCLate} + (1 - f_{Mo}^{NCLate})\, \varepsilon^{95}Mo^{CC} = \varepsilon^{95}Mo^{BSE} \quad (2)$$

where $\varepsilon^{94}Mo^{CC}$, $\varepsilon^{95}Mo^{CC}$, $\varepsilon^{94}Mo^{BSE}$ and $\varepsilon^{95}Mo^{BSE}$ are the average values of $\varepsilon^{94}Mo$ and $\varepsilon^{95}Mo$ for CC meteorites and BSE, respectively, and $a_{95Mo}$ and $b_{95Mo}$ are the coefficients of the NC line in the $\varepsilon^{95}Mo$–$\varepsilon^{94}Mo$ plane. This system of equation gives the solution:

$$\varepsilon^{94}Mo^{NCLate} = (b_{95Mo}\,\varepsilon^{94}Mo^{CC} - b_{95Mo}\,\varepsilon^{94}Mo^{BSE} - \varepsilon^{94}Mo^{CC}\,\varepsilon^{95}Mo^{BSE} + \varepsilon^{95}Mo^{CC}\,\varepsilon^{94}Mo^{BSE})/ (- a_{95Mo}\,\varepsilon^{94}Mo^{CC} + a_{95Mo}\,\varepsilon^{94}Mo^{BSE} - \varepsilon^{95}Mo^{BSE} + \varepsilon^{95}Mo^{CC}) \quad (3)$$

$$f_{Mo}^{NCLate} = (- a_{95Mo}\,\varepsilon^{94}Mo^{CC} + a_{95Mo}\,\varepsilon^{94}Mo^{BSE} - \varepsilon^{95}Mo^{BSE} + \varepsilon^{95}Mo^{CC})/( - a_{95Mo}\,\varepsilon^{94}Mo^{CC} - b_{95Mo} + \varepsilon^{95}Mo^{CC}) \quad (4)$$

To determine $\varepsilon^{94}Mo^{NCEarly}$, $f_{Bulk}^{CC}$, and $f_{Bulk}^{NCLate}$, let us start by considering two equations for the isotope anomalies of two lithophile elements, such as $\varepsilon^{54}Cr$ and $\varepsilon^{96}Zr$:

$$(a_{54Cr}\,\varepsilon^{94}Mo^{NCLate} + b_{54Cr})\, f_{Bulk}^{NCLate} + f_{Bulk}^{CC}\,\varepsilon^{54}Cr^{CC} + (1- f_{Bulk}^{NCLate} - f_{Bulk}^{CC})\,(a_{54Cr}\,\varepsilon^{94}Mo^{NCEarly} + b_{54Cr}) = \varepsilon^{54}Cr^{BSE} \quad (5)$$

$$(a_{96Zr}\,\varepsilon^{94}Mo^{NCLate} + b_{96Zr})\, f_{Bulk}^{NCLate} + f_{Bulk}^{CC}\,\varepsilon^{96}Zr^{CC} + (1- f_{Bulk}^{NCLate} - f_{Bulk}^{CC})\,(a_{96Zr}\,\varepsilon^{94}Mo^{NCEarly} + b_{96Zr}) = \varepsilon^{96}Zr^{BSE} \quad (6)$$

where the meaning of the variables is like before, just substituting $^{95}Mo$ with $^{54}Cr$ or $^{96}Zr$. These two equations have the same functional form, even if their coefficients are different. Thus, instead of giving two unknowns as function of the third, there is one value of $f_{Bulk}^{CC}$ that makes the two equations linearly dependent (that is, equivalent to each other). This is:

$$f_{Bulk}^{CC} = (a_{54Cr}\, b_{96Zr} - a_{54Cr}\,\varepsilon^{96}Zr^{BSE} - a_{96Zr} b_{54Cr} + a_{96Zr}\,\varepsilon^{54}Cr^{BSE})/(a_{54Cr}\, b_{96Zr} - a_{54Cr}\,\varepsilon^{96}Zr^{CC} - a_{96Zr} b_{54Cr} + a_{96Zr}\,\varepsilon^{54}Cr^{CC}) \quad (7)$$

In this equation $f_{Bulk}^{CC}$ represents the mass fraction of a lithophile element that was added to Earth by CC material. It can be safely assumed that this is the fraction of Earth's mass delivered by CC bodies. This assumption would not been valid only if most of the NC and CC bodies accreted by Earth (and Mars) were not chondritic in terms of the mass fraction of lithophile elements (i.e., because they were accreted as discrete cores or mantle debris from differentiated planetesimals). However, given that the Fe/Si ratio of the Earth is very close to chondritic, this can be excluded. For the value of $f_{Bulk}^{CC}$ given in (7), each of the equations (5) and (6) give the same relationship $\varepsilon^{94}Mo^{NCEarly}$ ($f_{Bulk}^{NCLate}$). Thus, it is not possible to determine $\varepsilon^{94}Mo^{NCEarly}$ and $f_{Bulk}^{NCLate}$ separately. Instead, for any other value of $f_{Bulk}^{CC}$ the equations (5) and (6) do not have any common intersection. Adding a third equation for another lithophile element (e.g., $\varepsilon^{145}Nd$, name it eq. 8), would not improve the situation since it would have the same functional form as (5) and (6). Ideally, the same value of $f_{Bulk}^{CC}$ would render the three equations equivalent to each other. In practice, given the errors in the various coefficients, the value of $f_{Bulk}^{CC}$ obtained by considering (5) and (6), (6) and (8), or (5) and (8) will be close, but not identical. Thus, we will only consider eq. (5) and (6) and use eq. (6) and (8) for control.

Because all coefficients in equations (1), (2), (5), and (6) have uncertainties, it does not make sense to use these equations to derive a single value for $\varepsilon^{94}Mo^{NCLate}$ , $f_{Mo}^{NCLate}$ and $f_{Bulk}^{CC}$. Thus, we used a Monte Carlo approach instead. In each trial, we pick a value for each coefficient within its Gaussian uncertainty, up to 3σ from its mean value. Note that correlated uncertainties were taken into account. The uncertainties of the a and b coefficients of each NC-line (say $a_{54Cr}$ and $b_{54Cr}$) are correlated because the line has to pass through the datapoints of NC meteorites. We impose that the NC-line in the $\varepsilon^{95}Mo$–$\varepsilon^{94}Mo$ plot passes through the point ($\varepsilon^{94}Mo$=0.5, $\varepsilon^{95}Mo$=0.206), that in the $\varepsilon^{54}Cr$–$\varepsilon^{94}Mo$ plot passes through ($\varepsilon^{94}Mo$=0.75, $\varepsilon^{54}Cr$=-0.483), that in the $\varepsilon^{96}Zr$–$\varepsilon^{94}Mo$ plot passes through ($\varepsilon^{94}Mo$=0.7, $\varepsilon^{96}Zr$=0.4), and that in the $\varepsilon^{145}Nd$–$\varepsilon^{94}Mo$ plot passes through ($\varepsilon^{94}Mo$=0.54, $\varepsilon^{145}Nd$=0.05). These points are chosen at the locations where the data constraining the NC-lines are the tightest. Another correlation exists between the value of $\varepsilon^{94}Mo^{CC}$ and $\varepsilon^{95}Mo^{CC}$, which is given by the CC line (*18, 51*). The mean values and the 1σ uncertainties of all independent quantities considered in the Monte Carlo calculation are provided in Supplementary Materials Table S4. Those concerning the isotopic anomalies of CC meteorites are computed from the mean and root mean square of the mean values of all types of CC meteorites (Supplementary Materials Data S1). In the Monte Carlo calculations, we performed 5 million trials. The results for $\varepsilon^{94}Mo^{NCLate}$ , $f_{Mo}^{NCLate}$ and $f_{Bulk}^{CC}$ are then binned in the intervals [-3:1], [0:1] and [0:1] and the resulting probability distribution functions are reported in Fig. 5 . As a control, we repeated the Monte Carlo calculation using the equation for $\varepsilon^{145}Nd$ instead of $\varepsilon^{96}Zr$. The resulting PDF for $f_{Bulk}^{CC}$ is very similar. Instead of having a maximum at $f_{Bulk}^{CC}$ =0.04 it monotonically decays from 0. Only 7.7% of the trials give $f_{Bulk}^{CC}$ >0.2 (instead of 2.5% in the case of Fig. 5).
We have also applied the same method for Mars. The values of the isotopic anomalies for bulk silicate Mars (BSM) are reported in the Supplementary Materials Table S5. The PDFs for the unknowns $\varepsilon^{94}Mo^{NCLate}$ , $f_{Mo}^{NCLate}$ and $f_{Bulk}^{CC}$ are shown in Fig. 5 . Our results show that the value of $f_{Mo}^{NCLate}$ for Mars is very similar to that for the Earth, indicating that the CC contributions to Mo in the mantles of both planets are comparable. Instead, the CC fraction to bulk Mars is much more restricted than that of the Earth, not exceeding 0.1. The value of $\varepsilon^{94}Mo^{NCLate}$ is larger for Mars than for the Earth, as can also be deduced graphically from Fig. 5, meaning–in our interpretation–that Mars did not reach material as close to the Sun as the Earth did. This makes sense given that Mars is 50% further away.

**Acknowledgments**
We thank NASA and the Meteorite Working Group, the UNM Meteorite Museum, and the Institut für Planetologie for providing samples for this study. Funded by the Deutsche Forschungsgemeinschaft (DFG, German Research Foundation)–Project-ID 263649064–TRR170. This is TRR 170 pub. no. 146.

**Data and materials availability:** All data needed to evaluate the conclusions in the paper are present in the paper and/or the Supplementary Materials.

**Supplementary Materials**
Figures S1 to S3
Tables S1 to S5
Data file S1

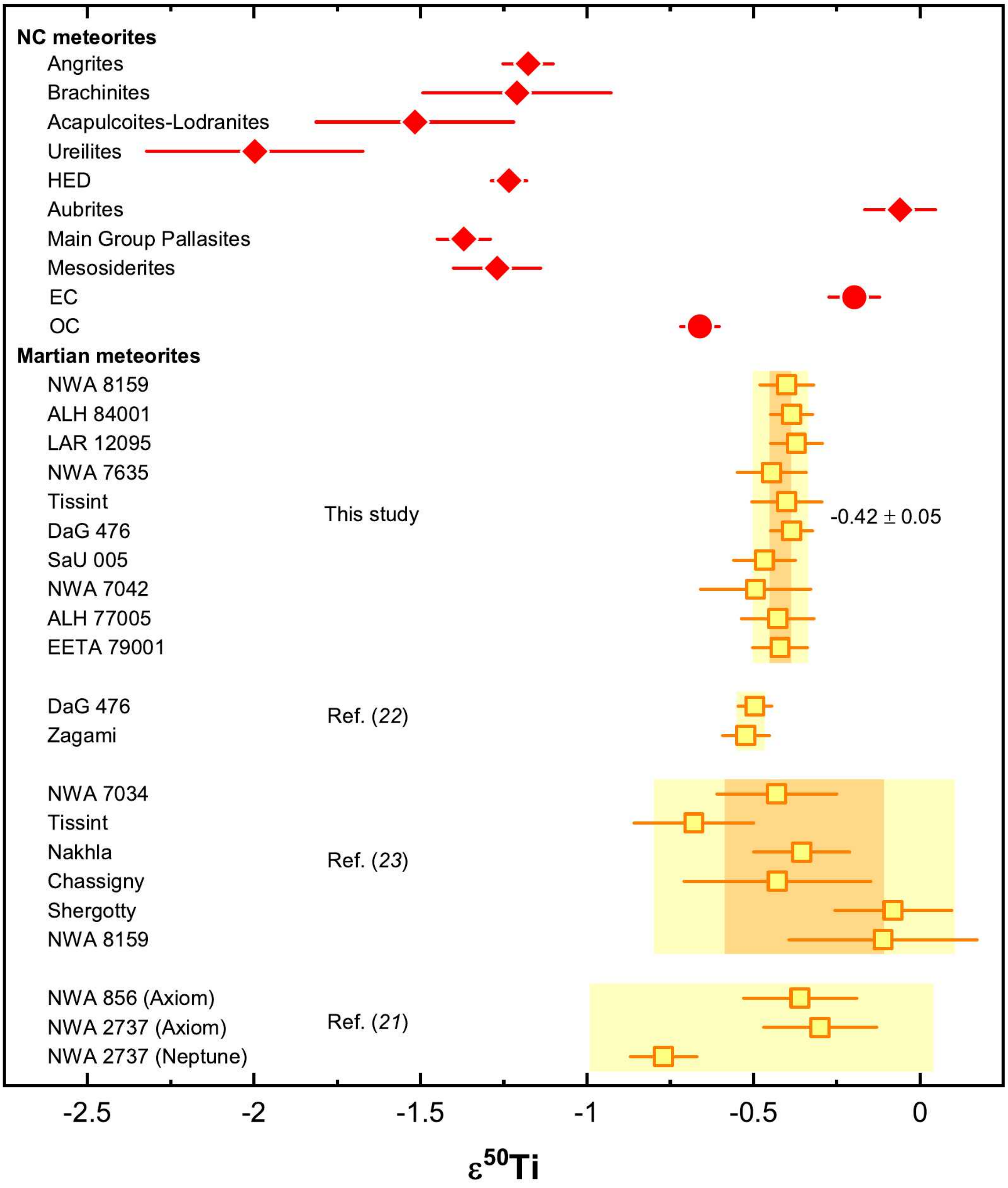

**Fig. S1. Isotope anomalies in $\varepsilon^{50}Ti$ for individual martian meteorites obtained here in comparison to literature data of martian meteorites and average NC bodies.** Data and data sources for NC meteorites are given in the Supplementary Materials Data S1. Individual error bars and orange band represent two-sided Student's t-values 95% confidence intervals, yellow band two times the standard deviation.

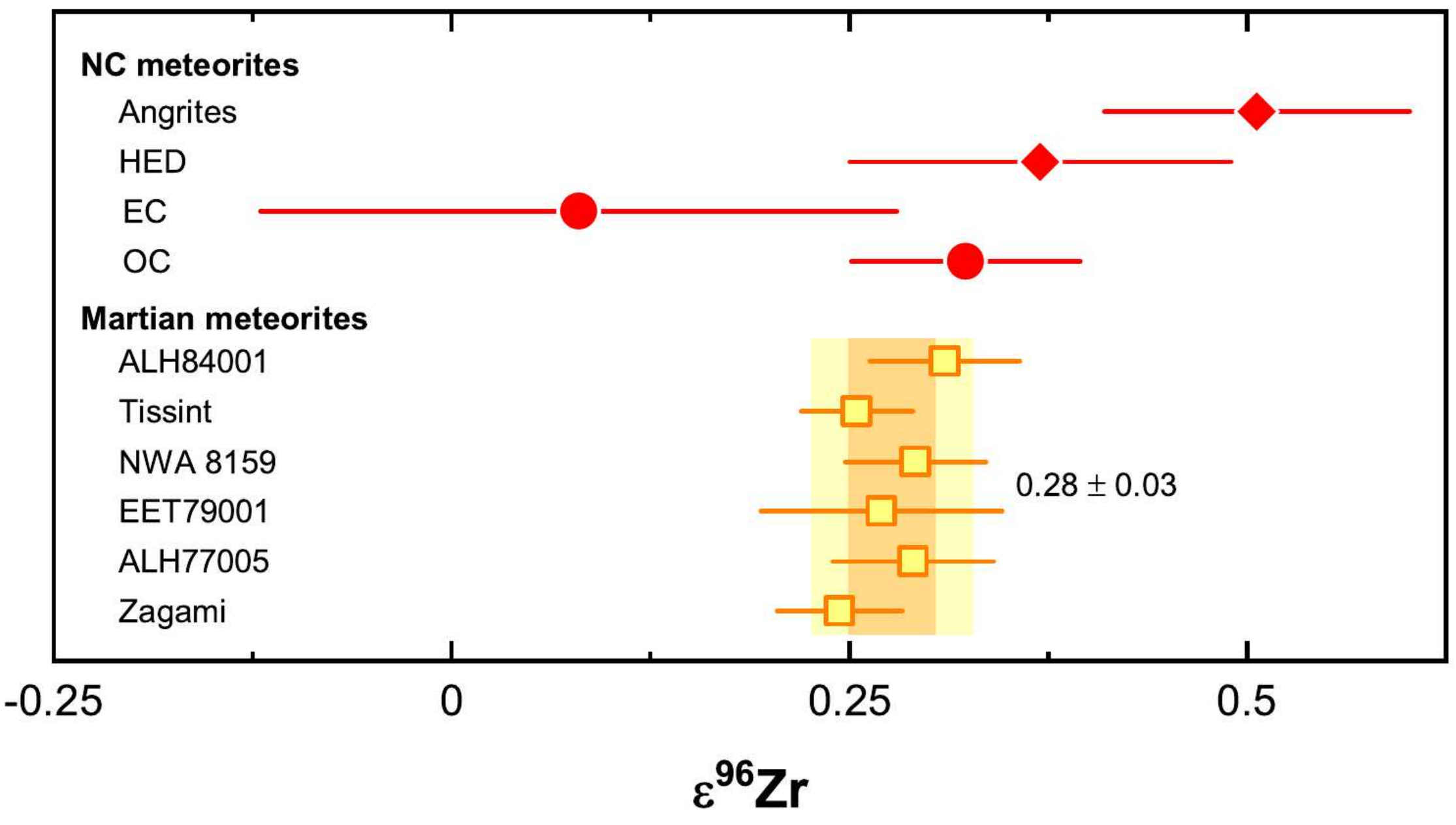

**Fig. S2. Anomalies in $\varepsilon^{96}$Zr for individual martian meteorites obtained here in comparison to literature data of NC bodies.** Data and data sources for NC meteorites are given in the Supplementary Materials Data S1. Individual error bars and orange band represent two-sided Student's t-values 95% confidence intervals, yellow band two times the standard deviation.

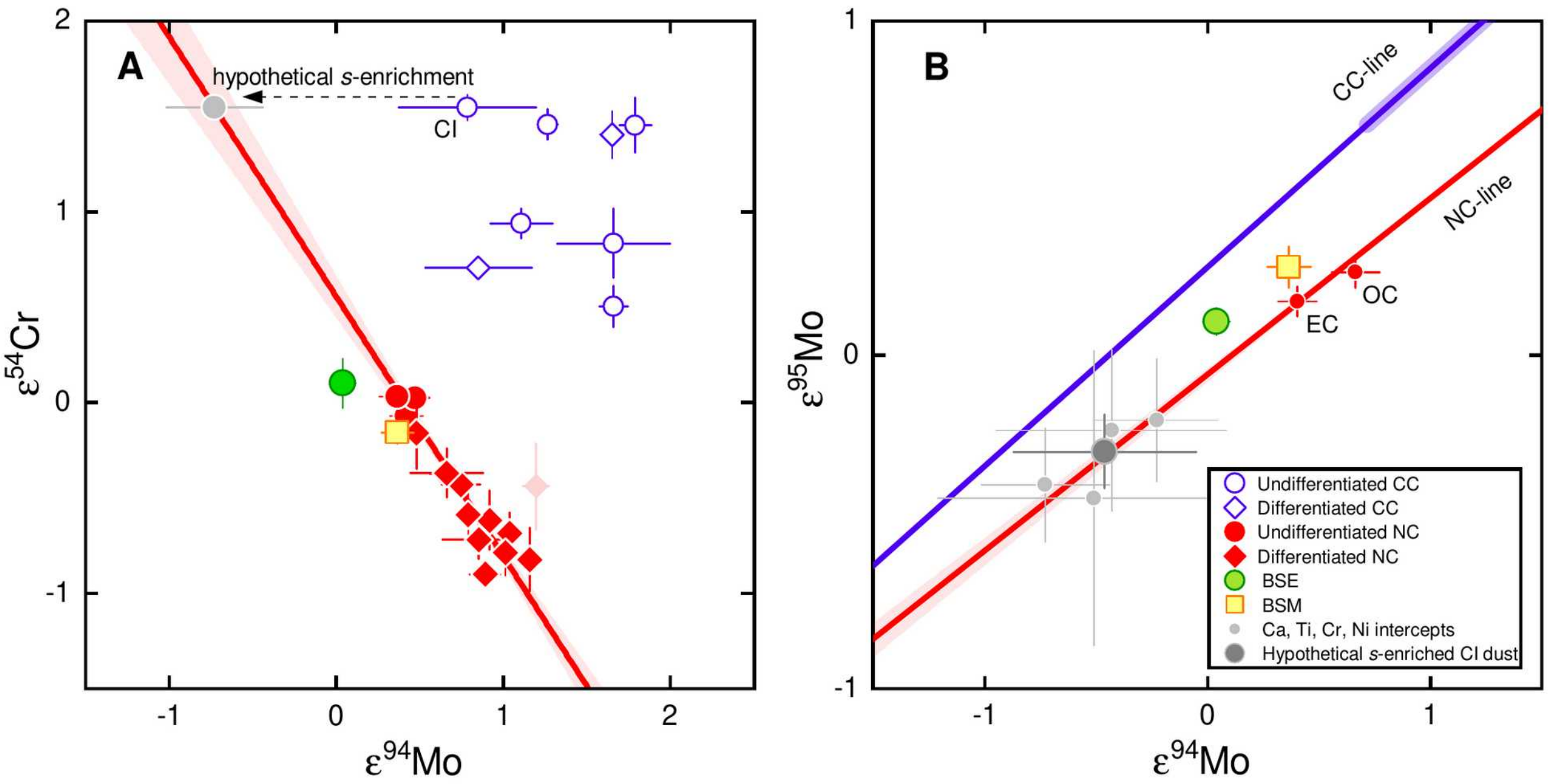

**Fig. S3. Position of putative *s*-process enriched CC dust in $\varepsilon^{54}$Cr-$\varepsilon^{94}$Mo and $\varepsilon^{95}$Mo-$\varepsilon^{94}$Mo isotope space.** (**A**) Extrapolation of the NC mixing trend to the intersection with the Ca, Ti, Cr, and Ni isotopic composition of CI meteorites allow to quantify the isotopic composition of a putative *s*-process-enriched CC dust component. When this is done using the NC correlations of $\varepsilon^{92}$Mo, $\varepsilon^{94}$Mo, $\varepsilon^{95}$Mo, $\varepsilon^{97}$Mo, and $\varepsilon^{100}$Mo with $\varepsilon^{48}$Ca, $\varepsilon^{50}$Ti, $\varepsilon^{54}$Cr, and $\varepsilon^{62}$Ni, respectively, the following Mo isotopic compositions for the putative *s*-enriched CC dust component are obtained: For a CI-like $\varepsilon^{48}$Ca of 2.11±0.20: $\varepsilon^{92}$Mo= -0.41±0.33, $\varepsilon^{94}$Mo= -0.43±0.29, $\varepsilon^{95}$Mo= -0.23±0.13, $\varepsilon^{97}$Mo= -0.35±0.11, $\varepsilon^{100}$Mo= 0.12±0.98. For a CI-like $\varepsilon^{50}$Ti of 1.89±0.15: $\varepsilon^{92}$Mo= -0.31±0.18, $\varepsilon^{94}$Mo= -0.23±0.14, $\varepsilon^{95}$Mo= -0.20±0.10, $\varepsilon^{97}$Mo= -0.02±0.05, $\varepsilon^{100}$Mo= 0.07±0.07. For a CI-like $\varepsilon^{54}$Cr of 1.55±0.07: $\varepsilon^{92}$Mo= -0.89±0.36, $\varepsilon^{94}$Mo= -0.73±0.29, $\varepsilon^{95}$Mo= -0.39±0.17, $\varepsilon^{97}$Mo= -0.26±0.14, $\varepsilon^{100}$Mo= -0.10±0.19. For a CI-like $\varepsilon^{62}$Ni of 0.17±0.09: $\varepsilon^{92}$Mo= -0.43±0.55, $\varepsilon^{94}$Mo= -0.51±0.55, $\varepsilon^{95}$Mo= -0.$^{43}$±0.33, $\varepsilon^{97}$Mo= -0.25±0.19, $\varepsilon^{100}$Mo= -0.16±0.16. All of these compositions (small gray points in **B**) are similar to one another, resulting in a weighted mean *s*-process Mo excess of $\varepsilon^{92}$Mo= -0.56±0.49, $\varepsilon^{94}$Mo= -0.46±0.41, $\varepsilon^{95}$Mo= -0.29±0.11, $\varepsilon^{97}$Mo= -0.18±0.24, $\varepsilon^{100}$Mo= -0.10±0.13 (large gray point in **B**). However, all these compositions plot on the NC-line close to the BSE in $\varepsilon^{95}$Mo-$\varepsilon^{94}$Mo isotope space, and not on the intersection of the NC and CC line, as would be expected for an internally consistent model. Thus, the missing component to explain the NC mixing trends cannot be inward drifting *s*-process enriched CC dust. Using the average isotopic composition of the CC bodies instead of CI values in the extrapolation does not change this conclusion. Uncertainties represent 95% confidence intervals.

**Table S1. Titanium isotopic composition of martian meteorites.**

| Sample | Type | N | $\varepsilon^{46}$Ti ± 95% ci | $\varepsilon^{48}$Ti ± 95% ci | $\varepsilon^{50}$Ti ± 95% ci |
|---|---|---|---|---|---|
| DaG476 | Depleted shergottite | 12 | -0.13 ± 0.06 | 0.02 ± 0.05 | -0.39 ± 0.06 |
| LAR12095 | Depleted shergottite | 12 | -0.12 ± 0.06 | 0.02 ± 0.12 | -0.37 ± 0.08 |
| NWA7635 | Depleted shergottite | 11 | -0.14 ± 0.11 | 0.04 ± 0.08 | -0.45 ± 0.10 |
| SaU005 | Depleted shergottite | 12 | -0.12 ± 0.13 | -0.01 ± 0.06 | -0.47 ± 0.09 |
| Tissint | Depleted shergottite | 7 | -0.01 ± 0.19 | 0.03 ± 0.09 | -0.40 ± 0.11 |
| ALH77005 | Intermediate shergottite | 12 | 0.01 ± 0.08 | 0.06 ± 0.15 | -0.43 ± 0.11 |
| EETA79001 | Intermediate shergottite | 12 | -0.16 ± 0.06 | 0.06 ± 0.04 | -0.42 ± 0.08 |
| NWA7042 | Intermediate shergottite | 7 | -0.11 ± 0.13 | -0.05 ± 0.11 | -0.49 ± 0.17 |
| ALH84001 | Orthopyroxenite | 11 | -0.13 ± 0.06 | 0.02 ± 0.05 | -0.39 ± 0.06 |
| NWA8159 | Augite basalt | 7 | -0.10 ± 0.09 | 0.03 ± 0.05 | -0.40 ± 0.08 |
| **Mars mean** | | **10** | **-0.10 ± 0.07** | **0.02 ± 0.04** | **-0.42 ± 0.05** |

N = number of measurements; $\varepsilon^{i}\mathrm{Ti} = [(^{i}\mathrm{Ti}/^{47}\mathrm{Ti})_{sample}/(^{i}\mathrm{Ti}/^{47}\mathrm{Ti})_{standard} - 1] \times 10^4$, where all ratios have been corrected for mass fractionation by internal normalization to fixed $^{49}\mathrm{Ti}/^{47}\mathrm{Ti}$ using the exponential law. Uncertainties represent two-sided Student's t-values 95% confidence intervals.

**Table S2. Zirconium isotopic composition of martian meteorites and terrestrial rock standards.**

| Sample | Type | N | $\varepsilon^{91}Zr$ ± 95% ci | $\varepsilon^{92}Zr$ ± 95% ci | $\varepsilon^{96}Zr$ ± 95% ci |
|---|---|---|---|---|---|
| Tissint | Depleted shergottite | 8 | -0.02 ± 0.03 | -0.04 ± 0.02 | 0.25 ± 0.04 |
| ALH77005 | Intermediate shergottite | 10 | 0.01 ± 0.01 | -0.02 ± 0.02 | 0.29 ± 0.06 |
| EET79001 | Intermediate shergottite | 9 | 0.01 ± 0.03 | 0.00 ± 0.01 | 0.27 ± 0.09 |
| Zagami | Enriched shergottite | 17 | -0.01 ± 0.02 | -0.01 ± 0.02 | 0.24 ± 0.04 |
| NWA8159 | Augite basalt | 8 | -0.02 ± 0.04 | -0.04 ± 0.02 | 0.29 ± 0.05 |
| ALH84001 | Orthopyroxenite | 13 | 0.00 ± 0.02 | -0.01 ± 0.02 | 0.31 ± 0.05 |
| **Mars mean** | | **6** | **0.00 ± 0.01** | **-0.02 ± 0.02** | **0.28 ± 0.03** |
| BHVO-2 | Hawaiian basalt standard | 18 | 0.01 ± 0.02 | 0.00 ± 0.02 | 0.01 ± 0.07 |
| BCR-2 | Intraplate basalt standard | 17 | -0.01 ± 0.02 | -0.01 ± 0.02 | -0.04 ± 0.04 |
| **terrestrial mean** | | **35** | **0.00 ± 0.01** | **0.00 ± 0.01** | **-0.01 ± 0.04** |

N= number of measurements; $\varepsilon^{i}Zr = [(^{i}Zr/^{90}Zr)sample/(^{i}Zr/^{90}Zr)standard - 1]\times 10^4$, where all ratios have been corrected for mass fractionation by internal normalization to fixed $^{94}Zr/^{90}Zr$ using the exponential law. Uncertainties represent two-sided Student's t-values 95% confidence intervals. BCR-2 data were obtained as part of the same measurement campaign, and have been reported in ref. (*47*).

**Table S3. Molybdenum isotopic composition of martian meteorites.**

| Sample | Type | N | $\varepsilon^{92}$Mo ± 2se | $\varepsilon^{94}$Mo ± 2se | $\varepsilon^{95}$Mo ± 2se | $\varepsilon^{97}$Mo ± 2se | $\varepsilon^{100}$Mo ± 2se |
|---|---|---|---|---|---|---|---|
| DaG476 | Depleted shergottite | 1 | 0.42 ± 0.22 | 0.34 ± 0.15 | 0.28 ± 0.12 | 0.16 ± 0.09 | -0.01 ± 0.13 |
| DaG476 | Depleted shergottite | 1 | 0.49 ± 0.21 | 0.33 ± 0.15 | 0.19 ± 0.12 | 0.14 ± 0.09 | -0.06 ± 0.13 |
| Tissint 1 | Depleted shergottite | 1 | 0.75 ± 0.21 | 0.62 ± 0.15 | 0.37 ± 0.11 | 0.14 ± 0.07 | -0.01 ± 0.12 |
| Tissint 2 | Depleted shergottite | 1 | 0.76 ± 0.22 | 0.42 ± 0.17 | 0.19 ± 0.12 | 0.07 ± 0.07 | 0.06 ± 0.15 |
| A (SaU005; DaG476) | Depleted shergottites | 1 | 0.49 ± 0.22 | 0.27 ± 0.16 | 0.19 ± 0.11 | 0.12 ± 0.08 | 0.08 ± 0.16 |
| B (ALH77005; EETA79001; NWA7042) | Intermediate shergottites | 1 | 0.55 ± 0.20 | 0.36 ± 0.14 | 0.34 ± 0.11 | 0.23 ± 0.09 | 0.08 ± 0.14 |
| C (NWA8159; RBT04262; NWA4864; LAR12011; LAR12095; ALH84001; NWA7635; Zagami; Nakhla; NWA10153; MIL03346) | Enriched shergottites; intermediate shergottites; depleted shergottites; Nakhlites; orthopyroxenite; augite basalt | 1 | 0.50 ± 0.23 | 0.24 ± 0.15 | 0.22 ± 0.12 | 0.16 ± 0.10 | 0.01 ± 0.15 |
| Tissint 2 and A-C (combined remains) | All types | 1 | 0.66 ± 0.22 | 0.35 ± 0.16 | 0.32 ± 0.11 | 0.13 ± 0.08 | 0.16 ± 0.13 |
| **Martian mantle mean** | | **8** | **0.58 ± 0.11** | **0.37 ± 0.10** | **0.26 ± 0.06** | **0.14 ± 0.04** | **0.04 ± 0.06** |

N= number of measurements; $\varepsilon^{i}\text{Mo} = [({}^{i}\text{Mo}/{}^{96}\text{Mo})_{\text{sample}}/({}^{i}\text{Mo}/{}^{96}\text{Mo})_{\text{standard}} - 1] \times 10^{4}$, where all ratios have been corrected for mass fractionation by internal normalization to fixed $^{98}$Mo/$^{96}$Mo using the exponential law. Due to low Mo content, some samples were combined to obtain enough Mo for a single measurement (samples A, B, C, and Tissint 2 and A-C combined remains). Uncertainties represent 2 s.e. for individual measurements and two-sided Student's t-values 95% confidence intervals for martian mantle average.

**Table S4. Mean values and the 1σ uncertainty of all independent quantities considered in the Monte Carlo calculation for the Earth.**

| coefficient | Mean | σ |
|---|---|---|
| $a_{95Mo}$ | 0.528 | 0.0225 |
| $a_{54Cr}$ | -1.456 | 0.145 |
| $a_{96Zr}$ | 1.2 | 0.485 |
| $a_{145Nd}$ | 0.1 | 0.035 |
| $\varepsilon^{94}Mo^{CC}$ | 1.65 | 0.855 |
| $\varepsilon^{54}Cr^{CC}$ | 1.14 | 0.305 |
| $\varepsilon^{96}Zr^{CC}$ | 0.77 | 0.275 |
| $\varepsilon^{145}Nd^{CC}$ | 0.03 | 0.025 |
| $\varepsilon^{94}Mo^{BSE}$ | 0.04 | 0.03 |
| $\varepsilon^{95}Mo^{BSE}$ | 0.1 | 0.02 |
| $\varepsilon^{54}Cr^{BSE}$ | 0.1 | 0.065 |
| $\varepsilon^{96}Zr^{BSE}$ | 0.01 | 0.015 |
| $\varepsilon^{145}Nd^{BSE}$ | -0.01 | 0.015 |

**Table S5. Mean values and the 1σ uncertainty of independent quantities considered in the Monte Carlo calculation for Mars.**

| coefficient | Mean | σ |
|---|---|---|
| $a_{95Mo}$ | 0.528 | 0.0225 |
| $a_{54Cr}$ | -1.456 | 0.145 |
| $a_{96Zr}$ | 1.2 | 0.485 |
| $\varepsilon^{94}Mo^{CC}$ | 1.65 | 0.855 |
| $\varepsilon^{54}Cr^{CC}$ | 1.14 | 0.305 |
| $\varepsilon^{96}Zr^{CC}$ | 0.77 | 0.275 |
| $\varepsilon^{94}Mo^{BSM}$ | 0.33 | 0.045 |
| $\varepsilon^{95}Mo^{BSM}$ | 0.25 | 0.045 |
| $\varepsilon^{54}Cr^{BSM}$ | -0.16 | 0.015 |
| $\varepsilon^{96}Zr^{BSM}$ | 0.28 | 0.01 |

**Table S1. Nucleosynthetic anomalies in planetary materials.**

| | Reservoir | $\varepsilon^{48}$Ca ± 95% CI | $\varepsilon^{50}$Ti ± 95% CI | $\varepsilon^{54}$Cr ± 95% CI | $\varepsilon^{54}$Fe ± 95% CI | $\varepsilon^{62}$Ni ± 95% CI | $\varepsilon^{96}$Zr ± 95% CI | $\varepsilon^{94}$Mo ± 95% CI | $\varepsilon^{95}$Mo ± 95% CI | $\varepsilon^{100}$Ru ± 95% CI | $\varepsilon^{145}$Nd ± 95% CI |
|---|---|---|---|---|---|---|---|---|---|---|---|
| Carbonaceous Chondrites | | | | | | | | | | | |
| CI | CC | 2.11 ± 0.20 | 1.89 ± 0.15 | 1.55 ± 0.07 | -0.02 ± 0.03 | 0.17 ± 0.09 | 0.34 ± 0.54 | 0.79 ± 0.41 | 0.69 ± 0.23 | -0.24 ± 0.13 | 0.02 ± 0.02 |
| CM | CC | 3.14 ± 0.14 | 2.89 ± 0.15 | 0.97 ± 0.08 | 0.23 ± 0.04 | 0.11 ± 0.03 | 0.70 ± 0.67 | 4.82 ± 0.20 | 3.17 ± 0.16 | -0.69 ± 0.38 | 0.06 ± 0.03 |
| CO | CC | 3.87 ± 0.56 | 3.58 ± 0.85 | 0.83 ± 0.18 | 0.13 ± 0.08 | 0.11 ± 0.04 | 0.94 ± 0.20 | 1.66 ± 0.34 | 1.39 ± 0.34 | -0.92 ± 0.98 | 0.06 ± 0.02 |
| CV | CC | 3.50 ± 0.38 | 3.45 ± 0.19 | 0.94 ± 0.08 | 0.22 ± 0.04 | 0.11 ± 0.03 | 0.96 ± 0.37 | 1.11 ± 0.19 | 0.93 ± 0.14 | -1.17 ± 0.22 | 0.04 ± 0.04 |
| CK | CC | | 3.42 ± 1.05 | 0.50 ± 0.11 | 0.26 ± 0.06 | | 0.45 ± 0.25 | 1.66 ± 0.09 | 1.30 ± 0.17 | -1.10 ± 0.23 | |
| CR | CC | 2.15 ± 0.22 | 2.51 ± 0.45 | 1.27 ± 0.06 | 0.29 ± 0.04 | 0.07 ± 0.08 | 1.03 ± 0.86 | 3.11 ± 0.15 | 2.26 ± 0.04 | -0.76 ± 0.36 | 0.04 ± 0.04 |
| CH | CC | | | 1.45 ± 0.14 | 0.16 ± 0.07 | | | 1.79 ± 0.10 | 1.29 ± 0.04 | -0.91 ± 0.13 | |
| CB | CC | | 2.04 ± 0.07 | 1.35 ± 0.20 | | 0.16 ± 0.11 | 0.96 ± 0.25 | 1.26 ± 0.04 | 0.99 ± 0.04 | -1.04 ± 0.04 | |
| CL | CC | | 2.60 ± 0.22 | 0.72 ± 0.08 | | | | | | | |
| ungrouped | | | | | | | | | | | |
| Tagish Lake | CC | 2.91 ± 0.05 | 2.76 ± 0.26 | 1.33 ± 0.26 | | | | | | -1.03 ± 0.13 | -0.01 ± 0.04 |
| NWA 1839 | CC | | 3.20 ± 0.51 | 1.03 ± 0.07 | | | | | | | |
| Flensburg | CC | | 2.98 ± 0.10 | 1.06 ± 0.11 | | | | | | | |
| EET 83226 | CC | | 4.25 ± 0.15 | 0.93 ± 0.13 | | | | | | | |
| EET 83355 | CC | | 3.11 ± 0.15 | 0.76 ± 0.13 | | | | | | | |
| MAC 87300 | CC | | 4.67 ± 0.15 | 0.71 ± 0.14 | | | | | | | |
| NWA 5958 | CC | | 3.34 ± 0.15 | 1.18 ± 0.16 | | | | | | | |
| LEW 85332 | CC | | 2.42 ± 0.15 | 1.23 ± 0.13 | | | | | | | |
| MAC 88107 | CC | | 3.03 ± 0.15 | 1.11 ± 0.15 | | | | | | | |
| MAC 87301 | CC | | 4.12 ± 0.15 | 0.83 ± 0.14 | | | | | | | |
| GRO 95566 | CC | | 3.50 ± 0.15 | 0.92 ± 0.13 | | | | | | | |
| LAP 04757 | NC | | -0.19 ± 0.15 | -0.33 ± 0.13 | | | | | | | |
| LAP 04773 | NC | | -0.54 ± 0.15 | -0.46 ± 0.16 | | | | | | | |
| Ordinary Chondrites | | | | | | | | | | | |
| H | NC | -0.23 ± 0.03 | -0.65 ± 0.14 | -0.31 ± 0.13 | 0.08 ± 0.03 | -0.06 ± 0.03 | 0.32 ± 0.08 | 0.72 ± 0.20 | 0.29 ± 0.05 | -0.27 ± 0.04 | 0.07 ± 0.04 |
| L | NC | -0.32 ± 0.11 | -0.67 ± 0.10 | -0.40 ± 0.06 | 0.09 ± 0.04 | -0.04 ± 0.04 | 0.40 ± 0.40 | 0.60 ± 0.19 | 0.21 ± 0.03 | -0.28 ± 0.13 | 0.04 ± 0.06 |
| LL | NC | -0.44 ± 0.09 | -0.67 ± 0.08 | -0.42 ± 0.08 | 0.13 ± 0.03 | -0.07 ± 0.03 | 0.34 ± 0.25 | 0.52 ± 0.10 | 0.18 ± 0.05 | -0.05 ± 0.13 | 0.01 ± 0.03 |
| OC Mean | NC | -0.31 ± 0.09 | -0.66 ± 0.06 | -0.37 ± 0.06 | 0.11 ± 0.03 | -0.06 ± 0.02 | 0.32 ± 0.04 | 0.67 ± 0.11 | 0.25 ± 0.05 | -0.24 ± 0.06 | 0.05 ± 0.03 |
| Rumuruti chondrites | | | | | | | | | | | |
| R | NC | | | -0.07 ± 0.03 | 0.06 ± 0.01 | | | 0.42 ± 0.10 | 0.18 ± 0.05 | -0.39 ± 0.13 | |
| Enstatite chondrites | | | | | | | | | | | |
| EH | NC | -0.32 ± 0.56 | -0.14 ± 0.07 | 0.02 ± 0.05 | | 0.03 ± 0.03 | 0.04 ± 0.24 | 0.47 ± 0.09 | 0.18 ± 0.07 | -0.08 ± 0.04 | 0.03 ± 0.04 |
| EL | NC | -0.40 ± 0.51 | -0.28 ± 0.17 | 0.03 ± 0.05 | 0.06 ± 0.01 | -0.03 ± 0.07 | 0.20 ± 0.40 | 0.36 ± 0.11 | 0.14 ± 0.06 | -0.08 ± 0.05 | 0.03 ± 0.02 |
| EC Mean | NC | -0.37 ± 0.37 | -0.20 ± 0.08 | 0.03 ± 0.03 | 0.06 ± 0.01 | 0.00 ± 0.03 | 0.08 ± 0.20 | 0.38 ± 0.08 | 0.15 ± 0.04 | -0.08 ± 0.03 | 0.03 ± 0.02 |
| Achondrites | | | | | | | | | | | |
| Acapulcoites-Lodranites | NC | | -1.52 ± 0.30 | -0.62 ± 0.15 | | | | 0.92 ± 0.07 | 0.48 ± 0.03 | -0.35 ± 0.05 | |
| Brachinites | NC | | -1.21 ± 0.28 | -0.44 ± 0.23 | | | | 1.20 ± 0.08 | 0.58 ± 0.05 | 0.25 ± 0.07 | |
| Winonaites | NC | -0.21 ± 0.09 | | | | | | 0.25 ± 0.15 | 0.09 ± 0.09 | -0.06 ± 0.10 | |
| Angrites | NC | -1.06 ± 0.33 | -1.18 ± 0.08 | -0.43 ± 0.06 | | 0.01 ± 0.05 | 0.51 ± 0.10 | 0.75 ± 0.11 | 0.39 ± 0.06 | | 0.07 ± 0.02 |
| Aubrites | NC | -0.44 ± 0.59 | -0.06 ± 0.11 | -0.16 ± 0.19 | | 0.05 ± 0.19 | | 0.48 ± 0.05 | 0.25 ± 0.06 | -0.06 ± 0.03 | 0.07 ± 0.07 |
| HED | NC | -1.24 ± 0.29 | -1.23 ± 0.05 | -0.69 ± 0.08 | 0.12 ± 0.02 | 0.03 ± 0.12 | 0.37 ± 0.12 | | | | 0.07 ± 0.01 |
| Ureilites | NC | -1.46 ± 0.20 | -2.00 ± 0.32 | -0.90 ± 0.04 | 0.14 ± 0.03 | -0.05 ± 0.16 | | 0.89 ± 0.09 | 0.38 ± 0.04 | -0.27 ± 0.11 | |
| ungrouped | | | | | | | | | | | |
| NWA 5363/5400 | NC | -0.53 ± 0.20 | -1.02 ± 0.10 | -0.37 ± 0.13 | | 0.01 ± 0.03 | | 0.66 ± 0.22 | 0.31 ± 0.15 | -0.34 ± 0.13 | 0.11 ± 0.06 |
| NWA 2526 | NC | | | | | | | 0.60 ± 0.13 | 0.39 ± 0.13 | -0.08 ± 0.13 | |
| NWA 6112 | NC | | | | | | | 1.55 ± 0.22 | 0.79 ± 0.15 | -0.46 ± 0.07 | |
| NWA 1058 | NC | | | | | | | 1.31 ± 0.11 | 0.68 ± 0.09 | -0.40 ± 0.07 | |
| NWA 8548 | NC | | | | | | | 1.53 ± 0.10 | 1.27 ± 0.07 | -1.14 ± 0.10 | |
| NWA 6926 | NC | | | | | | | 1.48 ± 0.12 | 1.14 ± 0.07 | -1.04 ± 0.14 | |
| NWA 7325 | NC | | -1.58 ± 0.33 | -0.61 ± 0.11 | | | | | | | |
| NWA 468 | NC | | -1.54 ± 0.42 | -0.59 ± 0.09 | | | | | | | |
| NWA 8054 | NC | | -1.01 ± 0.38 | -0.44 ± 0.08 | | | | | | | |
| GRV 020043 | NC | | -1.59 ± 0.24 | -0.48 ± 0.10 | | | | | | | |
| GRA 06128 | NC | | -1.44 ± 0.24 | -0.43 ± 0.11 | | | | | | | |
| GRA 06129 | NC | | -1.55 ± 0.27 | -0.46 ± 0.13 | | | | | | | |
| Bunburra Rockhole | NC | | | -0.36 ± 0.10 | | | 0.13 ± 0.03 | | | | 0.08 ± 0.04 |
| Tafassasset | CC | | 2.05 ± 0.10 | 1.40 ± 0.12 | | | | 1.65 ± 0.07 | 1.20 ± 0.05 | -1.15 ± 0.04 | 0.11 ± 0.02 |
| NWA 8548 | CC | | | | | | | 1.53 ± 0.10 | 1.27 ± 0.07 | -1.14 ± 0.10 | |
| NWA 6926 | CC | | | | | | | 1.48 ± 0.12 | 1.14 ± 0.07 | -1.04 ± 0.14 | |
| NWA 3100 | CC | | 1.91 ± 0.31 | 1.50 ± 0.11 | | | | | | | |
| NWA 2788 | CC | | 2.13 ± 0.51 | 1.04 ± 0.12 | | | | | | | |
| NWA 7822 | CC | | 2.13 ± 0.51 | 1.14 ± 0.08 | | | | | | | |
| NWA 2994 | CC | | 2.48 ± 0.96 | 1.31 ± 0.10 | | | | | | | |
| NWA 6704 | CC | | 2.07 ± 0.14 | 1.56 ± 0.10 | | | | | | | 0.09 ± 0.02 |
| Mesosiderites | NC | | -1.27 ± 0.13 | -0.69 ± 0.11 | | | | 1.04 ± 0.08 | 0.46 ± 0.05 | -0.42 ± 0.02 | |
| Pallasites | | | | | | | | | | | |
| Eagle station pallasites | CC | | | 0.71 ± 0.01 | 0.27 ± 0.42 | | | 0.85 ± 0.32 | 0.80 ± 0.14 | | |
| Main group pallasites | NC | | -1.37 ± 0.08 | -0.72 ± 0.10 | 0.09 ± 0.09 | -0.06 ± 0.10 | | 0.85 ± 0.22 | 0.38 ± 0.14 | -0.45 ± 0.23 | |
| ungrouped | | | | | | | | | | | |
| Milton | CC | | | 1.07 ± 0.07 | | | | 1.30 ± 0.26 | 1.04 ± 0.09 | -1.14 ± 0.15 | |
| Iron meteorites | | | | | | | | | | | |
| IAB | NC | | | | 0.00 ± 0.06 | -0.05 ± 0.06 | | 0.04 ± 0.10 | -0.07 ± 0.05 | 0.02 ± 0.09 | |
| IC | NC | | | | 0.06 ± 0.05 | -0.07 ± 0.04 | | 0.90 ± 0.06 | 0.40 ± 0.03 | -0.41 ± 0.05 | |
| IIAB | NC | | | -0.82 ± 0.17 | 0.12 ± 0.06 | -0.10 ± 0.07 | | 1.16 ± 0.04 | 0.53 ± 0.03 | -0.44 ± 0.05 | |
| IIC | CC | | | | 0.32 ± 0.03 | 0.16 ± 0.08 | | 2.31 ± 0.10 | 1.58 ± 0.05 | -1.03 ± 0.04 | |
| IID | CC | | | | | 0.19 ± 0.06 | | 1.18 ± 0.07 | 1.01 ± 0.03 | -0.86 ± 0.27 | |
| IIE | NC | | | -0.59 ± 0.13 | | | | 0.79 ± 0.05 | 0.36 ± 0.03 | | 0.04 ± 0.02 |
| IIF | CC | | | | | 0.09 ± 0.04 | | 1.10 ± 0.03 | 0.95 ± 0.04 | -1.02 ± 0.07 | |
| IIIAB | NC | | | -0.79 ± 0.12 | 0.10 ± 0.00 | -0.12 ± 0.02 | | 1.01 ± 0.04 | 0.46 ± 0.04 | -0.60 ± 0.06 | |
| IIIE | NC | | | | | -0.07 ± 0.04 | | 0.96 ± 0.02 | 0.46 ± 0.06 | -0.55 ± 0.06 | |
| IIIF | CC | | | | | 0.12 ± 0.08 | | 1.21 ± 0.05 | 0.97 ± 0.09 | -1.02 ± 0.11 | |
| IVA | NC | | | | | -0.07 ± 0.04 | | 0.79 ± 0.10 | 0.36 ± 0.05 | -0.29 ± 0.05 | |
| IVB | CC | | | | 0.29 ± 0.07 | 0.07 ± 0.04 | | 1.54 ± 0.10 | 1.16 ± 0.05 | -0.90 ± 0.05 | |
| ungrouped | | | | | | | | | | | |
| Wiley (IIC anomalous) | CC | | | | | 0.13 ± 0.05 | | 3.45 ± 0.10 | 2.24 ± 0.06 | -1.09 ± 0.08 | |
| South Byron Trio | CC | | | | | | | 1.27 ± 0.07 | 1.03 ± 0.04 | -1.07 ± 0.05 | |
| Mbosi | CC | | | | | | | 1.10 ± 0.43 | 1.02 ± 0.27 | | |
| Sombrerete (IAB complex) | CC | | | | | | | 1.73 ± 0.29 | 1.15 ± 0.15 | | |
| Tishomingo | CC | | | | | | | 1.42 ± 0.22 | 0.93 ± 0.13 | -1.00 ± 0.14 | |
| Chinga | CC | | | | | | | 1.62 ± 0.08 | 1.13 ± 0.02 | -1.04 ± 0.03 | |
| Dronino | CC | | | | | | | 1.39 ± 0.28 | 0.98 ± 0.15 | -1.02 ± 0.15 | |
| Mont Dieu | NC | | | | | | | 0.63 ± 0.21 | 0.18 ± 0.13 | | |
| Gebel Kamil | NC | | | | | | | 0.34 ± 0.30 | 0.07 ± 0.15 | -0.03 ± 0.06 | |
| Earth's mantle | BSE | 0.00 ± 0.04 | 0.00 ± 0.00 | 0.10 ± 0.13 | 0.00 ± 0.02 | 0.03 ± 0.02 | 0.01 ± 0.03 | 0.04 ± 0.06 | 0.10 ± 0.04 | 0.02 ± 0.02 | -0.01 ± 0.02 |
| Mars' mantle | BSM | -0.20 ± 0.03 | -0.42 ± 0.07 | -0.16 ± 0.03 | 0.07 ± 0.02 | 0.04 ± 0.03 | 0.28 ± 0.03 | 0.37 ± 0.10 | 0.26 ± 0.06 | | 0.02 ± 0.03 |
| CAIs | IC | 3.80 ± 1.33 | 8.57 ± 0.40 | 5.93 ± 0.52 | 0.41 ± 0.24 | 0.59 ± 0.17 | 1.76 ± 0.22 | 2.00 ± 1.55 | 2.31 ± 0.92 | -1.60 ± 0.06 | -0.22 ± 0.03 |

Uncertainties are Student-t 95% confidence intervals (CI), i.e. $(t_{0.95, N-1} \times s.d.)/\sqrt{N}$ for N≥4 and 2 s.d. for N<4. Data table is based on literature compilations (*20, 22, 53*), updated with recent publications (*9, 23, 42, 47, 52, 54–62*).